\pdfoutput=1
\RequirePackage{ifpdf}
\ifpdf 
\documentclass[pdftex]{sigma}
\else
\documentclass{sigma}
\fi

\numberwithin{equation}{section}

\newtheorem{Theorem}{Theorem}[section]
\newtheorem{Proposition}[Theorem]{Proposition}
\newtheorem{Conjecture}[Theorem]{Conjecture}
\newtheorem{Problem}[Theorem]{Problem}
 { \theoremstyle{definition}
\newtheorem{Definition}[Theorem]{Definition}
\newtheorem{Remark}[Theorem]{Remark} }

\DeclareMathOperator{\Li}{Li}

\begin{document}

\allowdisplaybreaks

\newcommand{\arXivNumber}{1712.09933}

\renewcommand{\thefootnote}{}

\renewcommand{\PaperNumber}{043}

\FirstPageHeading

\ShortArticleName{The Hyperbolic Asymptotics of Elliptic Hypergeometric Integrals}

\ArticleName{The Hyperbolic Asymptotics\\ of Elliptic Hypergeometric Integrals\\ Arising in Supersymmetric Gauge Theory\footnote{This paper is a~contribution to the Special Issue on Elliptic Hypergeometric Functions and Their Applications. The full collection is available at \href{https://www.emis.de/journals/SIGMA/EHF2017.html}{https://www.emis.de/journals/SIGMA/EHF2017.html}}}

\Author{Arash Arabi ARDEHALI}

\AuthorNameForHeading{A.A.~Ardehali}
\Address{School of Physics, Institute for Research in Fundamental Sciences (IPM),\\
P.O.~Box 19395-5531, Tehran, Iran}
\Email{\href{mailto:a.a.ardehali@gmail.com}{a.a.ardehali@gmail.com}}

\ArticleDates{Received January 09, 2018, in final form April 29, 2018; Published online May 06, 2018}

\Abstract{The purpose of this article is to demonstrate that $i)$ the framework of elliptic hypergeometric integrals (EHIs) can be extended by input from supersymmetric gauge theory, and $ii)$ analyzing the hyperbolic limit of the EHIs in the extended framework leads to a rich structure containing sharp mathematical problems of interest to supersymmetric quantum field theorists. Both of the above items have already been discussed in the
theoretical physics literature. Item $i$ was demonstrated by Dolan and Osborn in 2008. Item~$ii$ was discussed in the present author's Ph.D.~Thesis in 2016, wherein crucial elements were borrowed from the 2006 work of Rains on the hyperbolic limit of certain classes of EHIs. This article contains a concise review of these developments, along with minor refinements and clarifying remarks, written mainly for mathematicians interested in EHIs. In particular, we work with a representation-theoretic definition of a supersymmetric gauge theory, so that readers without any background in gauge theory~-- but familiar with the representation theory of semi-simple Lie algebras~-- can follow the discussion.}


\Classification{33D67; 33E05; 41A60; 81T13; 81T60}

\renewcommand{\thefootnote}{\arabic{footnote}}
\setcounter{footnote}{0}

\section{Introduction}

{\it Elliptic hypergeometric integrals} \cite{vanD:2000,vanD:2001,Spiridonov:2001,Spiridonov:2003} are multivariate (or matrix-) integrals of the form
\begin{gather*}
\mathcal{I}(p,q)=\int F(p,q;x_1,\dots,x_{r_G}) \, \mathrm{d}^{r_{G}}x,
\end{gather*}
taken over $-1/2\leq x_1,\dots,x_{r_G}\leq 1/2$, with $r_G$ the rank of some semi-simple matrix Lie group~$G$. The parameters $p$,~$q$ are often assumed to be complex numbers satisfying $0<|p|,|q|<1$, but for simplicity we take them in this article to be inside the open interval $]0,1[$ of the real line. The title ``elliptic hypergeometric integral'' (EHI) comes from the fact that the integrand $F$ involves ratios of products of a number of elliptic gamma functions. The definition of the elliptic gamma function can be found in equation~(\ref{eq:GammaDef}), and explicit EHIs can be found in Section~\ref{sec:review} below. For a very brief introduction to EHIs see \cite{Rains:talk}.

Extra complex parameters (denoted by $t_i$ in \cite{Spiridonov:2001}, for example) besides $p$, $q$ are often considered inside the arguments of the integrand $F$ and the EHI $\mathcal{I}$. Our EHIs here correspond to special cases where all those extra parameters are taken to be some powers of the product $pq$, such that their so-called ``balancing conditions''~-- as well as other constraints discussed in Section~\ref{sec:review}~-- are
satisfied. We will comment on those extra parameters briefly in Section~\ref{sec:review} and also in the appendix.

Mathematicians' interest in EHIs has been to a large extent due to the remarkable \emph{transformation identities} \cite{Rains:2005,Spiridonov:2001,Spiridonov:2003} of the form
\begin{gather}
\int F(p,q;x_1,\dots,x_{r_G}) \,\mathrm{d}^{r_{G}}x=\int \tilde{F}(p,q;y_1,\dots,y_{r_{\tilde{G}}})\, \mathrm{d}^{r_{\tilde{G}}}y,\label{eq:transInt}
\end{gather}
that they exhibit. We formally allow $r_{\tilde{G}}$ to be zero, in which case there is only a function of~$p$,~$q$~-- and no integral~-- on the right-hand side; then (\ref{eq:transInt}) would be an integral evaluation. In their magical flavor, the transformation identities of EHIs are somewhat analogous to, though generally much more non-trivial than, the celebrated Rogers--Ramanujan identities featuring in analytic number theory. A fruitful idea, beyond the scope of the present article, for studying EHI transformations has been the application of Bailey transforms to EHIs~\cite{Spiridonov:2004,SpiridonovWarnaar:2004}.

A major mathematical development in which EHIs played a key role is the elliptic generalization \cite{Rains:2006, Rains:2005} of the Koornwinder--Macdonald theory of orthogonal polynomials~\cite{Koornwinder:1992}. Specifically, in \cite{Rains:2006, Rains:2005} abelian biorthogonal functions were constructed whose biorthogonality relation is governed by the ``Type~II'' EHI of \cite{vanD:2000,vanD:2001,Spiridonov:2003}. This development, too, is beyond the scope of the present article, and the interested reader is encouraged to consult \cite{Rosengren:2017} for a better perspective.

Theoretical physicists' interest in EHIs started growing in 2008 when Dolan and Osborn \cite{Dolan:2008} showed that
\begin{itemize}\itemsep=0pt
\item four-dimensional supersymmetric (SUSY) gauge theory provides a~framework in which the classes of EHIs known at the time arise as a~particular partition function, called {\it the Romelsberger index}~\cite{Romelsberger:2005eg}, of some of the most famous models (namely SUSY QCD models with gauge group $G$ either unitary or symplectic);
\item the transformation identities of the EHIs have a very natural interpretation in the physical framework as the equality of the Romelsberger indices of a pair of electric-magnetic (or Seiberg-) dual models.
\end{itemize}
Since then, the physics community, often working together with mathematicians, started contributing to the mathematical theory by studying new EHIs arising in SUSY gauge theory, using SUSY dualities to conjecture new transformation identities, and sometimes also proving the new identities. References \cite{Kels:2017,Kutasov:2014,Spirido:2009,Spirido:2011or} are a few particularly clear demonstrations of the fruitfulness of this interplay between physics and mathematics. References \cite{Brunner:2017,Vartanov:2010} are examples of several works in the other direction, using rigorous mathematics to shed light on dualities in SUSY gauge theory. The relation between EHIs and SUSY gauge theory, and between the transformation identities and SUSY duality, is briefly reviewed in Section~\ref{sec:review} below.

The main focus of the present article is not the transformation properties of the EHIs though, but their rich asymptotic behavior in the so-called \emph{hyperbolic limit}, where $p,q\to 1$ while $\log p/\log q$ is kept fixed. Defining $b,\beta\in {}]0,\infty[$ through
\begin{gather*}
p=e^{-\beta b},\qquad q=e^{-\beta b^{-1}},
\end{gather*}
we have
\begin{gather}
\text{\emph{the hyperbolic limit:}}\quad \beta\to 0^+,\quad \text{with} \ b\in {} ]0,\infty[ \ \text{fixed}.\label{def:hypLim}
\end{gather}
The title ``hyperbolic'' comes from the fact that in this limit the elliptic gamma functions reduce to hyperbolic gamma functions; see Section~\ref{sec:defs} below for the details. EHIs also have nontrivial ``trigonometric'', ``rational'', and ``classical'' limits, which we do not consider here; the interested reader can learn more about these limits in \cite{Rains:2009}.

Mathematicians' interest in the hyperbolic limit of EHIs has been mainly because taking the hyperbolic limit of a transformation identity like (\ref{eq:transInt}), one often arrives at a reduction of the identity to the hyperbolic level:
\begin{gather*}
\int F_h(b;x_1,\dots,x_{r_G}) \, \mathrm{d}^{r_{G}}x=\int \tilde{F}_h(b;y_1,\dots,y_{r_{\tilde{G}}}) \, \mathrm{d}^{r_{\tilde{G}}}y,\label{eq:hypTransInt}
\end{gather*}
where now $F_h$ and $\tilde{F}_h$ involve hyperbolic (rather than elliptic) gamma functions, and the integrals range over $]{-}\infty,\infty[$.

While the study of the hyperbolic limit of EHIs goes back to \cite{vanD:2003,Stokman:2003}, rigorous asymptotic estimates were obtained first by Rains in 2006 \cite{Rains:2009} for certain special classes of EHIs. (See also \cite[Section~5]{vandeB:2007} where the results of \cite{Rains:2009} are used to further analyze the hyperbolic limit of certain EHIs.) In this article we review the work in \cite{Ardehali:2015c,Ardehali:thesis}, which used Rains's machinery to analyze the hyperbolic asymptotics of general EHIs arising in SUSY gauge theory. In Section~\ref{sec:hypLim} we first present the conjecture in~\cite{Ardehali:2015c} for the most general case, stating that
\begin{gather}
\mathcal I(b,\beta)\approx \left(\frac{2\pi}{\beta}\right)^{r_G}
\int \mathrm{d}^{r_G}x \, e^{-[\mathcal{E}^{\rm DK}_0(b,\beta)+V^{\mathrm{eff}}(\boldsymbol{x};b,\beta)] +i\Theta(\boldsymbol{x};\beta)},\label{eq:LagIndexSimp1Int}
\end{gather}
where $\approx$ means an $O\big(\beta^0\big)$ error after taking logarithms of the two sides. The symbol $\boldsymbol{x}$ denotes the collection $x_1,\dots,x_{r_G}$, and $\mathcal{E}^{\rm DK}_0$, $V^{\mathrm{eff}}$ are real functions of order $1/\beta$, while $\Theta$ is a real function of order $1/\beta^2$; see equations~(\ref{eq:EdkEquiv})--(\ref{eq:ThetaDef}) below for the explicit expressions. Next, we will specialize to the (still rather large) class of \emph{non-chiral} EHIs for which $\Theta=0$; this class encompasses all the EHIs studied by Rains~\cite{Rains:2009}. For non-chiral EHIs we present the precise analysis performed in~\cite{Ardehali:thesis}, and demonstrate that not only~(\ref{eq:LagIndexSimp1Int}) is true, but that in fact
\begin{gather}
\log\mathcal I(b,\beta)=-\big[\mathcal{E}^{\rm DK}_0(b,\beta)+V^{\mathrm{eff}}_{\min}(b,\beta)\big]+\dim \mathfrak{h}_{qu}\log\left(\frac{2\pi}{\beta}\right)+O\big(\beta^0\big),\label{eq:resultInt}
\end{gather}
where $\mathfrak{h}_{qu}$ is the locus of minima of $V^{\mathrm{eff}}(\boldsymbol{x};b,\beta)$ as a function of $\boldsymbol{x}$, and $V^{\mathrm{eff}}_{\min}(b,\beta)$ the value of $V^{\mathrm{eff}}(\boldsymbol{x};b,\beta)$ on this locus. (Recall that the generically leading terms $\mathcal{E}^{\rm DK}_0,V^{\mathrm{eff}}_{\min}$ are of order~$1/\beta$.) Finding the small-$\beta$ asymptotics of $\mathcal{I}(b,\beta)$ thus involves a minimization problem for $V^{\mathrm{eff}}$ as a~function of~$\boldsymbol{x}$. Interestingly, it turns out that $V^{\mathrm{eff}}$ is a piecewise linear function of $\boldsymbol{x}$; see the plots in Figs.~\ref{fig:A1}, \ref{fig:Sp4}, \ref{fig:SO5}, and~\ref{fig:ISS}.

We have not been able to prove general theorems on the minimum value or the dimension of the locus of minima of $V^{\mathrm{eff}}$, but have been able to address the minimization problem for specific EHIs, on a case-by-case basis, using Rains's generalized triangle inequalities~\cite{Rains:2009} or some variations thereof; see Section~\ref{sec:gti} for a few explicit examples.

The $O\big(\beta^0\big)$ term on the r.h.s.\ of (\ref{eq:resultInt}) is where the hyperbolic reduction of $\mathcal{I}(b,\beta)$ resides. We will not discuss this term in depth in the present article, and will only make brief remarks about it in certain examples in the last two sections; other examples for which this term is explicitly analyzed can be found in \cite{Ardehali:2015c,Rains:2009}.

Theoretical physicists' interest in the hyperbolic limit of EHIs has been partly because the hyperbolic reduction of the Romelsberger index $\mathcal{I}(b,\beta)$ of a $4d$ SUSY gauge theory\footnote{We follow the common terminology and refer to specific ``models'' in the gauge theory framework as ``theories''.} often yields the squashed three-sphere partition function $Z_{S^3}(b)$ of the dimensionally reduced~-- hence~$3d$~-- SUSY gauge theory; in other words the $O\big(\beta^0\big)$ term on the r.h.s.\ of (\ref{eq:resultInt}) is often $\log Z_{S^3}(b)$, with $Z_{S^3}(b)$ given in turn by a hyperbolic hypergeometric integral. This ties well with Rains's results for the hyperbolic reduction of the special classes of EHIs studied in \cite{Rains:2009}. The physics intuition for the reduction is roughly as follows. The index $\mathcal{I}(b,\beta)$ can be computed \cite{Assel:2014} by the path-integral of the SUSY gauge theory placed on Euclidean $S_b^3\times S_{\beta}^1$, where $S_b^3$ is the unit-radius squashed three-sphere with squashing parameter $b$, and $\beta$ is the circumference of the $S^1$. The $\beta\to 0$ limit shrinks $S^1_\beta$, hence leaving us with the dimensionally reduced theory on $S_b^3$. This reduction has been noticed quantitatively in some special cases \cite{Aharony:2013a,Dolan:2011sv,Gadde:2011ibn,Imamura:2011ibn,Niarchos:2012,Spiridonov:2012sv}. However, the mathematical results of \cite{Ardehali:2015c,Ardehali:thesis} clarify that the reduction works in the nice way encountered in \cite{Aharony:2013a,Dolan:2011sv,Gadde:2011ibn,Imamura:2011ibn,Niarchos:2012,Spiridonov:2012sv}~-- and so the above intuitive physical picture is correct~-- only when $V^{\mathrm{eff}}$ is minimized just at $\boldsymbol{x}=0$; this condition was satisfied in all the examples studied rigorously by Rains \cite{Rains:2009} as well. When this condition is \emph{not} satisfied, the hyperbolic reduction is more subtle. See Sections~\ref{subsec:soN} and \ref{subsec:ff} for two such more subtle examples.

There is an additional reason for the interest of the theoretical physics community in the hyperbolic limit of EHIs. Interpreting $S^1_\beta$ in $S_b^3\times S_{\beta}^1$ as the \emph{Euclidean time circle} of the background spacetime, we get an analogy with thermal quantum physics where the circumference~$\beta$ of the Euclidean time circle becomes the inverse temperature\footnote{The analogy with thermal physics is actually not quite precise. In the path-integral computation~\cite{Assel:2014} one must use a~supersymmetric (i.e., periodic) spin connection around $S^1_\beta$, whereas in thermal quantum physics the spin connection is anti-periodic around the Euclidean time circle. Nevertheless, as in \cite{Ardehali:2015c,Ardehali:thesis} we keep employing the analogy because it helps a useful import of intuition from thermal physics.}. The hyperbolic limit of the EHI then corresponds to the \emph{high-temperature} (or ``Cardy'') limit of the index $\mathcal{I}(b,\beta)$. Since the celebrated work of Cardy on the high-temperature asymptotics of 2d conformal field theory (CFT) partition functions \cite{Cardy:1986ie}, the ``Cardy asymptotics'' of various quantum field theory partition functions have been of interest in theoretical physics. In particular, for the special cases where the underlying SUSY gauge theory describes a~$4d$ CFT, the index $\mathcal{I}(b,\beta)$ encodes the analytic combinatorics of the supersymmetric operators in the CFT~\cite{Kinney:2005ej}. The hyperbolic asymptotics is then connected to the asymptotic degeneracy of the large-charge supersymmetric operators. The counting of these operators can then have implications, through the AdS/CFT correspondence \cite{Maldacena:1997h}, for heavy states of quantum gravity on anti-de~Sitter spacetimes \cite{Ardehali:2015c,Bourdier:2015,Kinney:2005ej}.

Because of this interest in the Cardy asymptotics of $\mathcal{I}(b,\beta)$, there had been other physical studies of the subject prior to \cite{Ardehali:2015c,Ardehali:thesis}. In particular, the leading Cardy asymptotics of $\mathcal{I}(b,\beta)$ was proposed in a well-known paper \cite{DiPietro:2014} to be given by $\log\mathcal I(b,\beta)\approx -\mathcal{E}^{\rm DK}_0(b,\beta)$. The mathematical results of \cite{Ardehali:2015c,Ardehali:thesis} clarified that this relation is modified, as in (\ref{eq:resultInt}), for non-chiral EHIs with $V^{\mathrm{eff}}_{\min}(b,\beta)\neq 0$. A physical understanding of this modification due to nonzero $V^{\mathrm{eff}}_{\min}$ is recently achieved in~\cite{DiPietro:2017}.

In the final section we mention some of the open problems of physical interest concerning the hyperbolic limit of EHIs.

\section{The required special functions and their asymptotics}\label{sec:defs}

\subsection*{The special functions and some of their useful properties}

For complex $a$, $q$ such that $0<|q|<1$, we define the \textit{Pochhammer symbol} as
\begin{gather*}
(a;q):=\prod_{k=0}^{\infty}\big(1-a q^k\big).
\end{gather*}
Often the notation $(a;q)_\infty$ is used for the above function; we are following Rains's convention~\cite{Rains:2009} of omitting the~$\infty$.

The Pochhammer symbol is related to the Dedekind eta function via
\begin{gather}
\eta(\tau)=q^{1/24}(q;q), \label{eq:etaPoc}
\end{gather}
with $q=e^{2\pi i\tau}.$ The eta function has an ${\rm SL}(2,\mathbb{Z})$ modular property that will be useful for us: $\eta(-1/\tau)=\sqrt{-i\tau}\eta(\tau)$.

The \textit{elliptic gamma function} (first introduced by Ruijsenaars in \cite{Ruijsenaars:1997}) can be defined for $\operatorname{Im}(\tau),\operatorname{Im}(\sigma) >0$ as
\begin{gather}
\Gamma(x;\sigma,\tau):=\prod_{j,k\ge 0}\frac{1-z^{-1}p^{j+1}q^{k+1}}{1-z p^{j}q^{k}},\label{eq:GammaDef}
\end{gather}
with $z:=e^{2\pi i x}$, $p:=e^{2\pi i \sigma}$, and $q:=e^{2\pi i \tau}$. The above expression gives a meromorphic function of $x,\sigma,\tau\in\mathbb{C}$. For generic choice of $\tau$ and $\sigma$, the elliptic gamma has simple poles at $x=l-m\sigma-n\tau$, with $m,n\in\mathbb{Z}_{\ge 0}$, $l\in\mathbb{Z}$.

We sometimes write $\Gamma(x;\sigma,\tau)$ as $\Gamma(z;p,q)$, or simply as $\Gamma(z)$. Also, the arguments of elliptic gamma functions are frequently written with ``ambiguous'' signs (as in $\Gamma(\pm x;\sigma,\tau)$); by that we mean a multiplication of several gamma functions each with a ``possible'' sign in the argument (as in $\Gamma(+x;\sigma,\tau)\times \Gamma(-x;\sigma,\tau)$). Similarly $\Gamma\big(z^{\pm
1}\big):=\Gamma(z;p,q)\times \Gamma\big(z^{-1};p,q\big)$.

The \textit{hyperbolic gamma function} (first introduced in a~slightly different form by Ruijsenaars in \cite{Ruijsenaars:1997})
can be defined, following Rains \cite{Rains:2009}, via
\begin{gather}
\Gamma_h(x;\omega_1,\omega_2):=\exp \left(\mathrm{PV}\int_{\mathbb{R}}\frac{e^{2\pi i x w}}{(e^{2\pi i\omega_1 w}-1)(e^{2\pi i\omega_2
w}-1)}\frac{\mathrm{d}w}{w}\right).\label{eq:hyperbolicGamma}
\end{gather}
The above expression makes sense only for $0<\operatorname{Im}(x)<2\operatorname{Im}(\omega)$, with $\omega:=(\omega_1+\omega_2)/2$. In that domain, the function defined by (\ref{eq:hyperbolicGamma}) satisfies
\begin{gather*}
\Gamma_h(x+\omega_2;\omega_1,\omega_2)=2\sin \left(\frac{\pi x}{\omega_1}\right)\Gamma_h(x;\omega_1,\omega_2),
\end{gather*}
which can then be used for an inductive meromorphic continuation of the hyperbolic gamma function to all $x\in\mathbb{C}$. For generic $\omega_1$, $\omega_2$ in the upper half plane, the resulting meromorphic function~$\Gamma_h(x;\omega_1,\omega_2)$ has simple zeros at $x=\omega_1\mathbb{Z}_{\ge1}+\omega_2\mathbb{Z}_{\ge1}$ and simple poles at $x=\omega_1\mathbb{Z}_{\le 0}+\omega_2\mathbb{Z}_{\le 0}$.

For convenience, we will frequently write $\Gamma_h(x)$ instead of $\Gamma_h(x;\omega_1,\omega_2)$, and $\Gamma_h(x\pm y)$ instead of
$\Gamma_h(x+y)\Gamma_h(x-y)$.

The hyperbolic gamma function has an important property that can be easily derived from the definition (\ref{eq:hyperbolicGamma}):
\begin{gather}
\Gamma_h(-\operatorname{Re}(x)+i\operatorname{Im}(x);\omega_1,\omega_2)=\big(\Gamma_h(\operatorname{Re}(x)
+i\operatorname{Im}(x);\omega_1,\omega_2)\big)^\ast,\label{eq:hyperbolicGammaConj}
\end{gather}
with $\ast$ denoting complex conjugation.

We also define the \textit{non-compact quantum dilogarithm} $\psi_b$ (cf.\ the function $e_b(x)$ in \cite{Faddeev:2001};
$\psi_b(x)=e_b(-i x)$) via
\begin{gather}
\psi_b(x):=e^{-i\pi x^2/2+i\pi(b^2+b^{-2})/24}\Gamma_h(ix+\omega;\omega_1,\omega_2),\label{eq:hyperbolicGammaPsi}
\end{gather}
where $\omega_1:=i b$, $\omega_2:=i b^{-1}$, and $\omega:=(\omega_1+\omega_2)/2$.
For generic choice of $b$, the zeros of $\psi_b(x)^{\pm 1}$ are of first order, and lie at $\pm\big(\big(b+b^{-1}\big)/2+b\mathbb{Z}_{\ge 0}+b^{-1}\mathbb{Z}_{\ge 0}\big)$. Upon setting $b=1$ we get the function~$\psi(x)$ of~\cite{Felder:1999}, i.e., $\psi_{b=1}(x)=\psi(x)$.

An identity due to Narukawa \cite{Narukawa:2004} implies the following important relation between $\psi_{b}(x)$ and the elliptic gamma function~\cite{Ardehali:2015b}:
\begin{gather}
\Gamma(x;\sigma,\tau)=\frac{e^{2i\pi Q_{-}(x;\sigma,\tau)}}{\psi_b\big(\frac{2\pi i x}{\beta}+\frac{b+b^{-1}}{2}\big)}\prod_{n=1}^{\infty}\frac{\psi_b\big({-}\frac{2\pi
in}{\beta}-\frac{2\pi i x}{\beta}-\frac{b+b^{-1}}{2}\big)}{\psi_b\big({-}\frac{2\pi in}{\beta}+\frac{2\pi i
x}{\beta}+\frac{b+b^{-1}}{2}\big)}\nonumber\\
\hphantom{\Gamma(x;\sigma,\tau)}{} =e^{2i\pi Q_{+}(x;\sigma,\tau)}\psi_b\left(-\frac{2\pi i
x}{\beta}-\frac{b+b^{-1}}{2} \right)\prod_{n=1}^{\infty}\frac{\psi_b\big({-}\frac{2\pi
in}{\beta}-\frac{2\pi i x}{\beta}-\frac{b+b^{-1}}{2}\big)}{\psi_b\big({-}\frac{2\pi in}{\beta}+\frac{2\pi i
x}{\beta}+\frac{b+b^{-1}}{2}\big)},\label{eq:GammaFVSq}
\end{gather}
where
\begin{gather*}
Q_{-}(x;\sigma,\tau) =-\frac{x^3}{6\tau\sigma}+\frac{\tau+\sigma-1}{4\tau\sigma}x^2-\frac{\tau^2+\sigma^2+3\tau\sigma-3\tau-3\sigma+1}{12\tau\sigma}x\\
\hphantom{Q_{-}(x;\sigma,\tau)=}{} -\frac{1}{24}(\tau+\sigma-1)\big(\tau^{-1}+\sigma^{-1}-1\big),\\
Q_{+}(x;\sigma,\tau)=Q_{-}(x;\sigma,\tau)+\frac{\big(x-\frac{\tau+\sigma}{2}\big)^2}{2\tau\sigma}-\frac{\tau^2+\sigma^2}{24\tau\sigma},
\end{gather*}
and
\begin{gather*}
\sigma=\frac{i\beta}{2\pi}b,\qquad \tau=\frac{i\beta}{2\pi}b^{-1}.
\end{gather*}
In the special case where $b=1$, the expressions in (\ref{eq:GammaFVSq}) are corollaries of Theorem~5.2 of~\cite{Felder:1999}. Therefore equation~(\ref{eq:GammaFVSq}) can be regarded as an extension of (and in fact was derived in~\cite{Ardehali:2015b} in an attempt to extend) Theorem~5.2 of~\cite{Felder:1999}.

\subsection*{The required asymptotic estimates}

Throughout this article we take the parameter $\beta$ to be real and strictly positive. Therefore by $\beta\to 0$ we always mean $\beta\to 0^+$.

We say $f(\beta)=O(g(\beta))$ as $\beta\to0$, if there exist positive real numbers $C,\beta_0$ such that for all $\beta<\beta_0$ we have $|f(\beta)|< C|g(\beta)|$. We say $f(x,\beta)=O(g(x,\beta))$ \emph{uniformly} over $S$ as $\beta\to0$, if there exist positive real numbers $C$, $\beta_0$ such that for all $\beta<\beta_0$ and all $x\in S$ we have $|f(x,\beta)|< C|g(x,\beta)|$.

We use the symbol $\sim$ when writing the \emph{all-orders asymptotics} of a function. For example, we have
\begin{gather*}
\log\big(\beta+e^{-1/\beta}\big)\sim \log\beta\qquad \text{as} \ \beta\to 0,
\end{gather*}
because we can write the left-hand side as the sum of $\log\beta$ and $\log\big(1+e^{-1/\beta}/\beta\big)$, and the latter is beyond all-orders in $\beta$.

More precisely, we say $f(\beta)\sim g(\beta)$ as $\beta\to 0$, if we have $f(\beta)- g(\beta)=O(\beta^n)$ for any (arbitrarily large) natural $n$.

The only unconventional piece of notation is the following: we will write $f(\beta)\simeq g(\beta)$ if $\log f(\beta)\sim \log g(\beta)$ (with an appropriate choice of branch for the logarithms). By writing $f(x,\beta)\simeq g(x,\beta)$ we mean that $\log f(x,\beta)\sim \log g(x,\beta)$ for all~$x$ on which $f(x,\beta),g(x,\beta)\neq 0$, and that $f(x,\beta)= g(x,\beta)=0$ for all $x$ on which either $f(x,\beta)=0$ or $g(x,\beta)=0$.

With the above notations at hand, we can asymptotically analyze the Pochhammer symbol as follows. The ``low-temperature'' $(T:=\beta^{-1}\to 0)$ behavior is trivial:
\begin{gather*}
\big(e^{-\beta};e^{-\beta}\big)\simeq 1\qquad \text{as} \ 1/\beta\to 0.
\end{gather*}

The ``high-temperature'' $(T^{-1}=\beta\to 0)$ asymptotics is nontrivial. It can be obtained using the ${\rm SL}(2,\mathbb{Z})$ modular property of the eta function, which yields
\begin{gather*}
\log \eta\left(\frac{i\beta}{2\pi}\right)\sim -\frac{\pi^2}{6\beta}+\frac{1}{2}\log\left(\frac{2\pi}{\beta}\right)\qquad \text{as} \ \beta\to 0.
\end{gather*}
The above relation, when combined with (\ref{eq:etaPoc}), implies
\begin{gather}
\log\big(e^{-\beta};e^{-\beta}\big)\sim -\frac{\pi^2}{6\beta}+\frac{1}{2}\log\left(\frac{2\pi}{\beta}\right)+\frac{\beta}{24}\qquad \text{as} \
\beta\to 0.\label{eq:PochAsy}
\end{gather}

For the hyperbolic gamma function, Corollary~2.3 of \cite{Rains:2009} implies that when $x\in\mathbb{R}$
\begin{gather}
\log\Gamma_h(x+r\omega;\omega_1,\omega_2)\sim -i\pi\big[x|x|/2 + (r-1)\omega|x|+(r-1)^2\omega^2/2+\big(b^2+b^{-2}\big)/24\big],\label{eq:hyperbolicGammaAsy}
\end{gather}
as $|x|\to\infty$, for any fixed real $r$, and fixed $b>0$.

Combining (\ref{eq:hyperbolicGammaAsy}) and (\ref{eq:hyperbolicGammaPsi}) we find that for fixed $\operatorname{Re}(x)$ and fixed $b>0$
\begin{gather*}
\log\psi_b(x)\sim 0\qquad \text{as} \ \beta\to 0, \quad \text{for} \ \operatorname{Im}(x)= -1/\beta 
\end{gather*}
with an exponentially small error, of the type $e^{-1/\beta}$.

The above estimate can be combined with (\ref{eq:GammaFVSq}) to yield the following estimates in the hyperbolic limit~(\ref{def:hypLim}):
\begin{gather}
\Gamma(x;\sigma,\tau) \simeq \begin{cases} \displaystyle \frac{e^{2i\pi
Q_{-}(x;\sigma,\tau)}}{\psi_b\big(\frac{2\pi i
x}{\beta}+\frac{b+b^{-1}}{2}\big)},& \text{for} \ -1<\operatorname{Re}(x)\le 0,\\
\displaystyle \simeq e^{2i\pi Q_{+}(x;\sigma,\tau)}\psi_b\left(-\frac{2\pi i
x}{\beta}-\frac{b+b^{-1}}{2}\right), & \text{for} \ 0\le\operatorname{Re}(x)<1, \end{cases} \label{eq:GammaAsyPosNegZ}
\end{gather}
with $\sigma=\frac{i\beta}{2\pi}b$, $\tau=\frac{i\beta}{2\pi}b^{-1}$, and with the range of $\operatorname{Re}(x)$ explaining our subscript
notations for~$Q_+$ and~$Q_-$. The relation (\ref{eq:GammaAsyPosNegZ}), combined with~(\ref{eq:hyperbolicGammaPsi}), demonstrates the reduction of the elliptic gamma function to the hyperbolic gamma function in the limit~(\ref{def:hypLim}).

As a result of (\ref{eq:GammaAsyPosNegZ}), for $x\in\mathbb{R}$ we have the following relations in the hyperbolic limit:
\begin{gather}
\Gamma\left(-x+\left(\frac{\tau+\sigma}{2}\right)r;\sigma,\tau\right) \simeq
\frac{e^{2i\pi Q_{-}(-\{x\}+(\frac{\tau+\sigma}{2})r;\sigma,\tau)}}{\psi_b\big({-}\frac{2\pi
i \{x\}}{\beta}-(r-1)\frac{b+b^{-1}}{2}\big)},\nonumber\\
\Gamma\left(x+\left(\frac{\tau+\sigma}{2}\right)r;\sigma,\tau\right) \simeq e^{2i\pi
Q_{+}(\{x\}+\left(\frac{\tau+\sigma}{2}\right)r;\sigma,\tau)}\psi_b\left(-\frac{2\pi
i \{x\}}{\beta}+(r-1)\frac{b+b^{-1}}{2} \right),\label{eq:GammaAsyPosNegZ2}
\end{gather}
with $\{x\}:=x-\lfloor x\rfloor$. The above estimates are first obtained in the range $0\le x<1$, and then extended to $x\in \mathbb{R}$ using the periodicity of the l.h.s.\ under $x\to x+1$.

\section[Elliptic hypergeometric integrals from supersymmetric gauge theory]{Elliptic hypergeometric integrals\\ from supersymmetric gauge theory}\label{sec:review}

\subsection{How a SUSY gauge theory with U(1) R-symmetry gives an EHI}\label{subsec:masterEHI}

For the purpose of the present article, we take the following essentially representation theoretic data to defines a $4d$ supersymmetric gauge theory with U(1) R-symmetry:
\begin{enumerate}\itemsep=0pt
\item[$i)$] a \emph{gauge group} $G$, which we take to be a semi-simple matrix Lie group of rank $r_G$, denote its root vectors by Dynkin labels $\alpha=(\alpha_1,\dots,\alpha_{r_G})$, while denoting the set of all the roots by $\Delta_G$;
\item[$ii)$] a finite number of \emph{chiral multiplets} $\chi_j$ (with $j=1,\dots,n_\chi$), to each of which we associate an \emph{R-charge} $r_j\in{}]0,2[$, and a~finite-dimensional irreducible representation $\mathcal{R}_j$ of~$G$, whose weight vectors we denote by $\rho^j:=(\rho^j_1,\dots,\rho^j_{r_G})$, while denoting the set of all the weights of $\mathcal{R}_j$ by $\Delta_{j}$.
\end{enumerate}

Note that even though we have as many $\alpha$s as $\dim G$ and as many~$\rho^j$s as $\dim \mathcal{R}_j$, we are not using further indices to label individual~$\alpha$s and~$\rho^j$s among these.

We further demand the following \emph{anomaly cancellation}, or ``consistency'', conditions:
\begin{subequations}\label{eq:anomalyCon}
\begin{gather}
\sum_j\sum_{\rho^j\in\Delta_{j}}\rho^j_l\rho^j_m\rho^j_n=0,\qquad \text{for all $l$, $m$, $n$},\label{eq:anomalyCon1}\\
\sum_j(r_j-1)\sum_{\rho^j\in\Delta_j}\rho^j_l\rho^j_m+\sum_{\alpha\in\Delta_G}\alpha_{l}\alpha_{m}=0, \qquad \text{for all $l$, $m$}. \label{eq:anomalyCon3}
\end{gather}
\end{subequations}

We can summarize our definition as follows.
\begin{Definition}\label{definition1} A {\it SUSY gauge theory with ${\rm U}(1)$ ${\rm R}$-symmetry} is a collection of the following data satisfying the relations~(\ref{eq:anomalyCon}): a semi-simple matrix Lie group $G$, and a finite number $n_\chi$ of pairs $\{\mathcal{R}_1, r_1\},\dots,\{\mathcal{R}_{n_\chi}, r_{n_\chi}\}$, where $\mathcal{R}_j$ are finite-dimensional irreducible representations of $G$ while $r_j$ are real numbers inside $]0,2[$. We denote the roots of $G$ by $\alpha$, the weights of $\mathcal{R}_j$ by $\rho^j$, and the set of all the weights of~$\mathcal{R}_j$ by~$\Delta_{j}$.
\end{Definition}

Although the field theory formulation of a SUSY gauge theory is beyond the scope of the present article, for the readers familiar with that formulation we add that
\begin{enumerate}\itemsep=0pt
\item[$i)$] the ``field content'' of a~SUSY gauge theory described as above is: a massless vector multiplet (containing the \emph{gauge field} and its fermionic super-partner \emph{gaugino fields}) transforming in the adjoint representation of $G$, a finite number $n_{\chi}$ of massless chiral multiplets (containing \emph{Weyl fermions} and their super-partner \emph{complex scalars}) transforming in $\mathcal{R}_j$ of $G$, and for each of the chiral multiplets a CP-conjugate multiplet, with R-charge $-r_j$, transforming in $\bar{\mathcal{R}_j}$;
\item[$ii)$] the constraints~(\ref{eq:anomalyCon}) are respectively the conditions for cancellation of the gauge$^3$ and U(1)$_R$-gauge$^2$ anomalies of the field theory\footnote{The gauge-${\rm U}(1)_R ^2$ and gauge-gravity$^2$ anomalies vanish automatically because we are focusing on semi-simple gauge groups (cf.\ equation~(\ref{eq:sumRho=0}) below); upon extending the framework to compact $G$, their cancellation should be demanded as extra consistency conditions besides~(\ref{eq:anomalyCon}). The gauge$^2$-gravity anomalies cancel between CP-conjugate Weyl fermions.};
\item[$iii)$] since in field theory one should also specify the \emph{interactions} of various fields, it would be more precise from that perspective to say that Definition~\ref{definition1} does not single out a unique SUSY gauge theory, but describes a universality class of SUSY gauge theories compatible with the specified data.
\end{enumerate}

Our next definition bridges SUSY gauge theory to EHIs, through the Romelsberger index \cite{Kinney:2005ej,Romelsberger:2005eg} (also referred to as ``the $4d$ supersymmetric index'' \cite{Rastelli:2016}, or ``the $4d$ superconformal index'' when applied to superconformal field theories~\cite{Kinney:2005ej}).
\begin{Definition}\label{definition2}
The {\it Romelsberger index} of a SUSY gauge theory with U(1) R-symmetry (as in Definition~\ref{definition1}) is given by
\begin{gather}
\boxed{\mathcal{I}(b,\beta):=\frac{(p;p)^{r_G}(q;q)^{r_G}}{|W|}\int
\left(\prod_{k=1}^{r_G}\frac{\mathrm{d}z_k}{2\pi i z_k}\right)
\frac{\prod\limits_j \prod\limits_{\rho^j \in\Delta_j}\Gamma\big((pq)^{r_j/2}
z^{\rho^j}\big)}{\prod\limits_{\alpha_+\in\Delta_G}\Gamma(
z^{\pm\alpha_+})}.}\label{eq:LagEquivIndex}
\end{gather}
Here, $p=e^{-\beta b}$, $q=e^{-\beta b^{-1}}$, and we take $\beta,b\in {}]0,\infty[$, so $p,q$ are real numbers in $]0,1[$. Our symbolic notation $z^{\rho^j}$ should be understood as $z_1^{\rho^j_1}\times\dots\times z_{r_G}^{\rho^j_{r_G}}$. The $\alpha_+$ are the positive roots of $G$, and $|W|$ is the order of the Weyl group of $G$. The integral is over the unit torus in the space of $z_k$, or alternatively over $x_k\in [-1/2,1/2]$ for $x_k$ defined through $z_k=e^{2\pi i x_k}$. By $z^{\alpha}$ we mean $z_1^{\alpha_1}\times\dots\times z_{r_G}^{\alpha_{r_G}}$.
\end{Definition}

In our notation the three-dimensional representation of ${\rm SU}(3)$, for example, has weights $(\rho_1,\rho_2)=(1,0),(0,1),(-1,-1)$, and the positive roots of ${\rm SU}(3)$ are $\alpha_+=(1,-1),(2,1),(1,2)$. Also, the parameters $b$, $\beta$ are related to Rains's parameters in~\cite{Rains:2009} via $\omega_1=ib$, $\omega_2=ib^{-1}$, and $v=\frac{\beta}{2\pi}$.

The expression (\ref{eq:LagEquivIndex}) is regarded in the physics literature as the outcome of a combinatorial (Hamiltonian)~\cite{Ardehali:thesis,Dolan:2008} or a path-integral (Lagrangian)~\cite{Assel:2014} computation, the starting point being more physical definitions for the Romelsberger index in both cases. For our purposes here though, it is more convenient to take (\ref{eq:LagEquivIndex}) as the definition. See~\cite{Rastelli:2016} for a recent review of the index from a more physical perspective.

One of the simplest examples of SUSY gauge theories with U(1) R-symmetry is the SU(2) supersymmetric QCD (SQCD) with three flavors: the gauge group~$G$ is SU(2); there are three ``quark'' chiral multiplets with R-charge~$1/3$ in the fundamental representation of SU(2), so $\mathcal{R}_1=\mathcal{R}_2=\mathcal{R}_3=\square$ and $\rho^{1}_1,\rho^{2}_1,\rho^{3}_1=\pm 1$; there are also three ``anti-quark'' chiral multiplets with R-charge $1/3$ in the anti-fundamental representation of SU(2), so $\mathcal{R}_4=\mathcal{R}_5=\mathcal{R}_6=\bar{\square}$ and $\rho^{{4}}_1,\rho^{{5}}_1,\rho^{{6}}_1=\mp1$. Since the positive root of SU(2) is $\alpha_+=2$, and its Weyl group has order~2, the expression~(\ref{eq:LagEquivIndex}) ends up being in this case
\begin{gather}
\mathcal{I}_{N_c=2,N_f=3}(b,\beta)=\frac{(p;p)(q;q)}{2}\int_{-1/2}^{1/2} \mathrm{d}x \frac{\Gamma^6\big((pq)^{1/6} z^{\pm 1} \big)}{\Gamma\big( z^{\pm 2}\big)}.\label{eq:LagIndexSU2SQCD}
\end{gather}
This is a special case of the elliptic beta integral of Spiridonov~\cite{Spiridonov:2001}, the first of the species of EHIs to have been discovered.

In the previous sentence we said ``a special case'', because, as alluded to in the introduction, EHIs often depend on extra parameters~($t_i$ in~\cite{Spiridonov:2001}, for example). We are focusing for simplicity on special cases where all these parameters are taken to be powers of~$pq$, such that their ``ba\-lan\-cing conditions'', as well as the constraints~(\ref{eq:anomalyCon}) following from their expression as in~(\ref{eq:LagEquivIndex}), are satisfied. Introducing those parameters back corresponds to turning on \emph{flavor fugacities}~-- or \emph{flavor chemical potentials}~-- in the physical picture. We briefly comment on the incorporation of flavor fugacities in the appendix.

Dolan and Osborn \cite{Dolan:2008} realized that the Romelsberger index of ${\rm SU}(N_c)$ SQCD with $N_f$ flavors (which has gauge group ${\rm SU}(N_c)$, and has $2N_f$ chiral multiplets of R-charge $1-N_c/N_f$, half of them in the fundamental and the other half in the anti-fundamental representation of the gauge group) corresponds to the EHI denoted $I^{(m)}_{A_n}$ in~\cite{Rains:2005}, with $n=N_c-1$ and $m=N_f-N_c-1$. The ${\rm Sp}(2N)$ gauge theory with $2N_f$ chiral multiplets of R-charge $1-(N+1)/N_f$ in the $2N$ dimensional fundamental representation gives rise to the EHI denoted $I^{(m)}_{BC_n}$ in~\cite{Rains:2005}, with $n=N$ and $m=N_f-N-2$. This is enough reason to claim that the expression~(\ref{eq:LagEquivIndex}) provides a~legitimate extension of the framework of EHIs. In summary, \textit{every supersymmetric gauge theory with a ${\rm U}(1)$ ${\rm R}$-symmetry} defined as above, \textit{gives} what may be called \textit{an elliptic hypergeometric integral}.

For brevity, we sometimes drop the adjective ``with ${\rm U}(1)$ R-symmetry'', but by a SUSY gauge theory we mean a SUSY gauge theory with U(1) R-symmetry throughout this article; the latter is the appropriate framework for EHIs, as explained above.

The general expression (\ref{eq:LagEquivIndex}) appears for instance in \cite{Assel:2014}. There, the constraints~(\ref{eq:anomalyCon}) were assumed, but the condition $0<r_j<2$ was not.

\begin{Remark}\label{remark1} The assumption $0<r_j$ guarantees that the poles of the gamma functions in the integrand of the EHI~(\ref{eq:LagEquivIndex}) are avoided, so $\mathcal{I}(b,\beta)$ is a continuous real function in the domain $b,\beta\in{}]0,\infty[$.
\end{Remark}
That $\mathcal{I}(b,\beta)$ is real follows from dividing the integral to two pieces, one over $x_1\in[-1/2,0]$, $x_{i>1}\in [-1/2,1/2]$, the other over $x_1\in[0,1/2]$, $x_{i>1}\in [-1/2,1/2]$, and then arguing that the two pieces are complex conjugates of each other because under $\boldsymbol{x}\to -\boldsymbol{x}$ the integrand goes to its complex conjugate. That $\mathcal{I}(b,\beta)$ is continuous on $b,\beta\in{}]0,\infty[$ follows from the continuity of the integrand when $0<r_j$.

The further constraint $r_j<2$ is imposed to make $\mathcal{I}(b,\beta)$ still better-behaved.
\begin{Remark}\label{remark2} The assumption $0<r_j<2$ allows using the estimates
\begin{gather*}
\frac{1}{\Gamma(z)}\simeq 1-z,\qquad\text{and}\qquad \Gamma\big((pq)^{r_j/2}z\big)\simeq 1,\label{eq:GammaLowT}
\end{gather*} as $1/\beta\to 0$, for fixed $x\in\mathbb{R}$ and fixed $b\in {}]0,\infty[$, both valid uniformly over $x\in\mathbb{R}$, so that we get a~universal ``low-temperature'' asymptotics for the Romelsberger index~(\ref{eq:LagEquivIndex}):
\begin{gather}
\mathcal{I}(b,\beta)\simeq\frac{1}{|W|}\int \mathrm{d}^{r_G}x
\prod_{\alpha_+}\big(\big(1- z^{\alpha_+}\big)\big(1- z^{-\alpha_+}\big)\big)=1\label{eq:WeylIntegral}
\end{gather} as $1/\beta\to 0$, for fixed $b\in {}]0,\infty[$.
The equality on the r.h.s.\ results from the Weyl integral formula. Such asymptotics are expected for partition functions of gapped quantum systems, whose only state contributing significantly to the partition function at low-enough temperatures is the vacuum state having unit Boltzmann factor. So the asymptotics~(\ref{eq:WeylIntegral}) is a nice property for the EHI to have.
\end{Remark}

We mention in passing that despite the anomaly cancellation conditions (\ref{eq:anomalyCon}), we may still have non-zero \emph{'t~Hooft anomalies}, which do not lead to inconsistencies or R-symmetry violations in the quantum gauge theory. A careful discussion of such 't~Hooft anomalies is beyond
the scope of the present article; the interested reader is referred to~\cite{Spiridonov:2012sv}.

\subsection{How SUSY dualities lead to transformation identities for EHIs}

The ${\rm SU}(2)$ SQCD theory with three flavors, whose index appeared in~(\ref{eq:LagIndexSU2SQCD}), has a magnetic (or Seiberg-) dual description as a theory of 15 chiral multiplets with R-charge $2/3$ without a gauge group (hence $r_{\tilde{G}}=0$). Equality of the indices computed from the two descriptions implies
\begin{gather*}
\frac{(p;p)(q;q)}{2}\int_{-1/2}^{1/2} \mathrm{d}x\, \frac{\Gamma^6\big((pq)^{1/6} z^{\pm 1} \big)}{\Gamma\big( z^{\pm
2}\big)}=\Gamma^{15}\big((pq)^{1/3}\big).
\end{gather*}
This is a special case (with flavor fugacities suppressed) of Spiridonov's elliptic beta integral formula~\cite{Spiridonov:2001}, the first of the EHI transformation identities to have been discovered.

The ${\rm SU}(N_c)$ SQCD theory with $N_f$ flavors described above, has a~Seiberg dual description with $\tilde{G}={\rm SU}(N_f-N_c)$, with $N_f$ magnetic quark chiral multiplets in the fundamental of~$\tilde{G}$ along with~$N_f$ magnetic anti-quark chiral multiplets in the anti-fundamental of~$\tilde{G}$, and $N_f^2$ magnetic ``mesons'' in the trivial representation of $\tilde{G}$; the magnetic quark and anti-quark multiplets have R-charge $N_c/N_f$, while the magnetic mesons have R-charge $2(1-N_c/N_f)$. The equality of the indices computed from the two descriptions implies the transformation identity~\cite{Rains:2005}
\begin{gather*}
I^{(m)}_{A_n} \big((pq)^{(m+1)/2(m+n+2)};(pq)^{(m+1)/2(m+n+2)};p,q\big)\\
\qquad{}= \Gamma\big((pq)^{(m+1)/(m+n+2)}\big)^{(m+n+2)^2} \cdot
I^{(n)}_{A_m}\big((pq)^{(n+1)/2(m+n+2)};(pq)^{(n+1)/2(m+n+2)};p,q\big).
\end{gather*}
Again, note that for simplicity we are suppressing flavor fugacities; in the language of~\cite{Rains:2005} we are focusing on the special case where all $t_i$ and $u_i$ are set equal to each other, hence~-- from their balancing condition~-- equal to~$(pq)^{(n+1)/2(m+n+2)}$.

Similarly, the transformation identity for $I^{(m)}_{BC_n}(p,q)$ can be arrived at from the SUSY duality for the corresponding ${\rm Sp}(2N)$ theory mentioned above. These and similar instances of the relation between SUSY dualities and EHI transformation were discovered in~\cite{Dolan:2008}. See~\cite{Spirido:2009,Spirido:2011or} for a more thorough discussion of these matters.

We would like to emphasize that currently no systematic procedure is known for deriving the electric-magnetic dual of a given SUSY gauge theory with U(1) R-symmetry. The most systematic method available is to write down the Romelsberger index of the given gauge theory, and to hope that somewhere in the mathematics literature there is a~transformation identity discovered for it; from the transformation identity one then reads the field content of the magnetic dual theory. Of course, achieving a general systematic procedure for \emph{transforming the master EHI}~(\ref{eq:LagEquivIndex}) would change the story. Although it might be too much to hope for, such a~procedure would allow for the first time a systematic approach to the derivation of electric-magnetic dualities between SUSY gauge theories with U(1) R-symmetry.

\section{The rich structure in the hyperbolic limit}\label{sec:hypLim}

We now attempt to understand the asymptotics of the master EHI (\ref{eq:LagEquivIndex}) in the hyperbolic limit: $\beta\to 0^+$ with $b>0$ fixed.

\subsection{A conjecture for the general case}

\begin{figure}[t]\centering
 \includegraphics[scale=1]{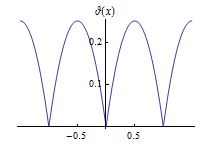} \qquad \includegraphics[scale=1]{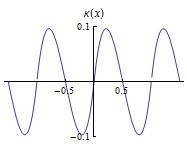}
\caption{The even, piecewise quadratic function $\vartheta(x)$ (on the left) and the odd, piecewise cubic func\-tion~$\kappa(x)$ (on the right). Both are continuous and periodic, and their fundamental domain can be taken to be $[-1/2,1/2]$.}\label{fig:thetakappa}
\end{figure}

The following \emph{uniform} estimate over $x\in\mathbb{R}$ \cite{Ardehali:2015c} can be used for a preliminary investigation of the $\beta\to0$ limit of (\ref{eq:LagEquivIndex}) (cf.\ \cite[Proposition~2.12]{Rains:2009}):
\begin{gather}
\log\Gamma\big((pq)^{r/2}z\big)=
i\frac{2\pi^3}{3\beta^2}\kappa(x)
+\frac{2\pi^2}{\beta}\left(\frac{b+b^{-1}}{2}\right)(r-1)\vartheta(x)\nonumber\\
\hphantom{\log\Gamma\big((pq)^{r/2}z\big)=}{} -\frac{\pi^2}{3\beta}\left(\frac{b+b^{-1}}{2}\right)(r-1)+O\big(\beta^0\big) \qquad
\text{for} \ r\in {}]0,2[ .\label{eq:GammaOffCenter2}
\end{gather}
As in \cite{Rains:2009}, we have defined the continuous, positive, even, periodic function
\begin{gather}
\vartheta(x):=\{x\}(1-\{x\}) \qquad \big(=|x|-x^2 \qquad\text{for} \ x\in[-1,1]\big).\label{eq:varthetaDef}
\end{gather}
We have also introduced the continuous, odd, periodic function
\begin{gather}
\kappa(x):=\{x\}(1-\{x\})(1-2\{x\}) \qquad \big(=2x^3-3x|x|+x \qquad \text{for} \ x\in[-1,1]\big).\label{eq:kappaDef}
\end{gather}
These functions are displayed in Fig.~\ref{fig:thetakappa}.

The estimate (\ref{eq:GammaOffCenter2}) can be derived from the second line of (\ref{eq:GammaFVSq}), but we need the following fact: for fixed $r\in{}]0,2[$ and fixed $b>0$, as $\beta\to 0$ the function $\log\psi_b\big({-}\frac{2\pi i \{x\}}{\beta}+(r-1)\frac{b+b^{-1}}{2}\big)$ is uniformly bounded over ($x\in)$ $\mathbb{R}$. It suffices of course to establish this fact in the ``fundamental domain'' $x\in [0,1[$. To obtain the uniform bound, divide this interval into $[0,N_0\beta]$ and $[N_0\beta,1[$, with $N_0$ chosen as follows. Since $\psi_b\big({-}2\pi i N+(r-1)\frac{b+b^{-1}}{2}\big)\to 1$ as $N\to\infty$, there is a large enough~$N_0$, so that for all $N>N_0$ we have $\psi_b\big({-}2\pi i N+(r-1)\frac{b+b^{-1}}{2}\big)\approx 1$, with an error of say~$.1$. With this choice of $N_0$ it is clear that $\log\psi_b\big({-}\frac{2\pi i x}{\beta}+(r-1)\frac{b+b^{-1}}{2}\big)$ is uniformly bounded over $[N_0\beta,1[$ (for all $\beta$ smaller than $1/N_0$). On the other hand, since $\log\psi_b\big({-}2\pi i x+(r-1)\frac{b+b^{-1}}{2}\big)$ is continuous, it is guaranteed to be uniformly bounded on the compact domain $[0,N_0]$; re-scaling $x\to\frac{x}{\beta}$ this implies the uniform bound on $\log\psi_b\big({-}\frac{2\pi i x}{\beta}+(r-1)\frac{b+b^{-1}}{2}\big)$ over $[0,N_0\beta]$, and we are done. Note that for $\log\psi_b\big({-}\frac{2\pi i \{x\}}{\beta}+(r-1)\frac{b+b^{-1}}{2}\big)$ to not diverge at $x\in\mathbb{Z}$, we need $r\big(\frac{b+b^{-1}}{2}\big)\notin b\mathbb{Z}_{\le 0}+b^{-1}\mathbb{Z}_{\le 0}$ and $(r-2)\big(\frac{b+b^{-1}}{2}\big)\notin b\mathbb{Z}_{\ge 0}+b^{-1}\mathbb{Z}_{\ge 0}$; our constraint $r\in{}]0,2[$ takes care of these.

The estimate (\ref{eq:GammaOffCenter2}) can not, however, be applied to the elliptic gamma functions in the denominator of (\ref{eq:LagEquivIndex}); these would require an analog of~(\ref{eq:GammaOffCenter2}) which would apply when $r=0$. This analog, which is valid uniformly over compact subsets of $\mathbb{R}$ avoiding an~$O(\beta)$ neighborhood of~$\mathbb{Z}$, reads (cf.\ \cite[Proposition~2.12]{Rains:2009})
\begin{gather}
\log\left(\frac{1}{\Gamma(z^{\pm 1})}\right)=
\frac{4\pi^2}{\beta}\left(\frac{b+b^{-1}}{2}\right)\vartheta(x)-\frac{2\pi^2}{3\beta}\left(\frac{b+b^{-1}}{2}\right)+O\big(\beta^0\big). \label{eq:GammaOffCenter2Vec}
\end{gather}
Note that the above relation would follow from a (sloppy) use of~(\ref{eq:GammaOffCenter2}) with $r=0$, but unlike (\ref{eq:GammaOffCenter2}) the above estimate is not valid uniformly over $\mathbb{R}$. For real $x$ in the dangerous neighborhoods of size $O(\beta)$ around~$\mathbb{Z}$, the following slightly weaker version of~(\ref{eq:GammaOffCenter2Vec}) applies (cf.\ \cite[Corollary~3.1]{Rains:2009}):
\begin{gather}
\frac{1}{\Gamma(z^{\pm 1})}=O\left(\exp
\left[\frac{4\pi^2}{\beta}\left(\frac{b+b^{-1}}{2}\right)\vartheta(x)-\frac{2\pi^2}{3\beta}\left(\frac{b+b^{-1}}{2}\right)\right]\right)\qquad \text{as} \ \beta\to 0.\label{eq:GammaOffCenter2VecB}
\end{gather}
A stronger estimate in this region can be obtained by relating the product on the l.h.s.\ to a~product of theta functions, and then using the modular property of theta functions. The weaker estimate above suffices for our purposes though.

Let us recall the asymptotic relation (\ref{eq:PochAsy}) from Section~\ref{sec:defs}:
\begin{gather*}
\log\big(e^{-\beta};e^{-\beta}\big)\sim -\frac{\pi^2}{6\beta}+\frac{1}{2}\log\left(\frac{2\pi}{\beta}\right)+\frac{\beta}{24}\qquad \text{as} \ \beta\to 0,
\end{gather*}
where $\sim$ indicates asymptotic equality to all orders. Combining this with (\ref{eq:GammaOffCenter2}) and (\ref{eq:GammaOffCenter2Vec}), a~sloppy simplification of the EHI in~(\ref{eq:LagEquivIndex}) now yields
\begin{gather}
\mathcal I(b,\beta)\approx \left(\frac{2\pi}{\beta}\right)^{r_G}
\int_{\mathfrak{h}_{\rm cl}} \mathrm{d}^{r_G}x\,
e^{-[\mathcal{E}^{\rm DK}_0(b,\beta)+V^{\mathrm{eff}}(\boldsymbol{x};b,\beta)]+i\Theta(\boldsymbol{x};\beta)}.\label{eq:LagIndexSimp1}
\end{gather}
We have denoted the unit hypercube $x_i\in [-1/2,1/2]$ by $\mathfrak{h}_{\rm cl}$, because in the path-integral picture the range of integration can be interpreted as the ``classical'' moduli-space of the gauge field holonomies~-- aka ``Wilson loops''~-- around~$S^1_\beta$. We have also defined
\begin{gather}
\mathcal{E}^{\rm DK}_0(b,\beta):=\frac{\pi^2}{3\beta}\left(\frac{b+b^{-1}}{2}\right)\left[\dim G+\sum_j(r_j-1)\dim \mathcal{R}_j\right],\label{eq:EdkEquiv}\\
V^{\mathrm{eff}}(\boldsymbol{x};b,\beta):=\frac{4\pi^2}{\beta}\left(\frac{b+b^{-1}}{2}\right)L_h(\boldsymbol{x}),\label{eq:VeffEquiv1}\\
\Theta(\boldsymbol{x};\beta):=\frac{8\pi^3}{\beta^2}Q_h(\boldsymbol{x}).\label{eq:ThetaDef}
\end{gather}
The \emph{real} functions $Q_h(\boldsymbol{x})$ and $L_h(\boldsymbol{x})$ are defined by
\begin{gather}
Q_h(\boldsymbol{x}):=\frac{1}{12}\sum_j\sum_{\rho^j\in\Delta_j}\kappa(\langle\rho^j\cdot \boldsymbol{x}\rangle),\nonumber\\ 
L_h(\boldsymbol{x}):= \frac{1}{2}\sum_j(1-r_j)\sum_{\rho^j \in\Delta_j}\vartheta(\langle\rho^j\cdot \boldsymbol{x}\rangle)-\sum_{\alpha_+\in\Delta_G}\vartheta(\langle\alpha_+\cdot \boldsymbol{x}\rangle),\label{eq:LhDef}
\end{gather}
where $\langle\ \cdot\ \rangle$ denotes the dot product.

Note that in (\ref{eq:LagIndexSimp1}) we are claiming that \emph{the EHI is approximated well with the integral of its approximate integrand}. This is far from obvious: while the estimate~(\ref{eq:GammaOffCenter2}) for the gamma functions in the numerator is valid uniformly over the domain of integration, the estimate (\ref{eq:GammaOffCenter2Vec}) for the denominator gamma functions is uniform only over compact subsets of~$\mathfrak{h}_{\rm cl}$ that avoid an~$O(\beta)$ neighborhood of the \emph{Stiefel diagram}
\begin{gather*}
\mathcal{S}_g=:\bigcup_{\alpha_+}\{\boldsymbol{x}\in \mathfrak{h}_{\rm cl}|\langle\alpha_+\cdot \boldsymbol{x}\rangle\in\mathbb{Z}\}.
\end{gather*}
Let's denote this neighborhood by $\mathcal{S}^{(\beta)}_g$. Intuitively speaking, we expect the estimate (\ref{eq:GammaOffCenter2VecB}), which applies also on $\mathcal{S}^{(\beta)}_g$, to guarantee that our unreliable use of (\ref{eq:GammaOffCenter2Vec}) over this small region modifies the asymptotics at most by a multiplicative $O\big(\beta^0\big)$ factor\footnote{A stronger version of (\ref{eq:GammaOffCenter2VecB}) implies that the expression (\ref{eq:LhDef}) for $L_h$ should be corrected on $\mathcal{S}^{(\beta)}_g$. However, the correction is of order one only in an $O(e^{-1/\beta})$ neighborhood of~$\mathcal{S}_g$. In particular, the corrected $L_h$ diverges on $\mathcal{S}_g$, because the integrand of the EHI vanishes there.}; in absence of unforseen cancellations due to integration, this factor is not $O(\beta)$. Since the errors of the estimates used in deriving (\ref{eq:LagIndexSimp1}) from (\ref{eq:LagEquivIndex}) are also multiplicative $O\big(\beta^0\big)$, we may hope the two sides of the symbol $\approx$ in (\ref{eq:LagIndexSimp1}) to be equal, asymptotically as $\beta\to 0$, up to a multiplicative factor of order~$\beta^0$ (but not order~$\beta$). For non-chiral theories with $\Theta=Q_h=0$ this intuitive argument can be made more precise, as we do below. We leave the validity of~(\ref{eq:LagIndexSimp1}) for the general case, with possibly nonzero~$Q_h$, as a~conjecture.

\begin{Conjecture}\label{conjecture}
For a general Romelsberger index $\mathcal{I}(b,\beta)$ as in Definition~{\rm \ref{definition2}}, the esti\-ma\-te~\eqref{eq:LagIndexSimp1} is valid, asymptotically in the hyperbolic limit, up to an $O\big(\beta^0\big)$ error upon taking logarithms of the two sides.
\end{Conjecture}

Studying the small-$\beta$ behavior of the multiple-integral on the r.h.s.\ of (\ref{eq:LagIndexSimp1}) is now a (rather nontrivial) exercise in standard asymptotic analysis. We have not been able to carry this analysis forward for the general case with $\Theta\neq0$, so we will shortly restrict attention to non-chiral theories which have $\Theta=0$. But before that, we comment on some important properties of the func\-tions~$Q_h$ and $L_h$ introduced above.

The real function $Q_h$ appearing in the phase $\Theta(\boldsymbol{x};\beta)$ is \emph{piecewise quadratic}, because the cubic terms in it cancel thanks to the anomaly cancellation condition~(\ref{eq:anomalyCon1}):
\begin{gather*}
\frac{\partial^3 Q_h(\boldsymbol{x})}{\partial x_l\partial x_m\partial x_n}=\sum_j\sum_{\rho^j\in\Delta_j}\rho^j_l\rho^j_m\rho^j_n=0.
\end{gather*}
Moreover, we have the identity
\begin{gather}
\sum_{\rho^j\in\Delta_j}\rho^j_l=0,\label{eq:sumRho=0}
\end{gather}
which follows from considering the action of a Weyl reflection, with respect to the hyperplane perpendicular to some simple root $\alpha_s\in\Delta_G$, on $\big\langle \sum\limits_{\rho^j\in\Delta_j}\rho^j\cdot\alpha_s\big\rangle$; the reflection only permutes the weights $\rho^j$, but negates~$\alpha_s$, and the completeness of the simple roots $\alpha_s$ as a~basis for the weight space establishes~(\ref{eq:sumRho=0}). Then we learn that $Q_h$ is stationary at the origin:
\begin{gather*}
\frac{\partial Q_h(\boldsymbol{x})}{\partial x_l}|_{\boldsymbol{x}=0}=\frac{1}{12}\sum_j\sum_{\rho^j\in\Delta_j}\rho^j_l=0.
\end{gather*}
It is easy to verify that $Q_h(\boldsymbol{x})$ has a continuous first derivative. Also, $Q_h(\boldsymbol{x})$ is odd under \smash{$\boldsymbol{x}\to -\boldsymbol{x}$}, and vanishes at $\boldsymbol{x}=0$; these properties follow from the fact that the function~$\kappa(x)$ defined in~(\ref{eq:kappaDef}) is a continuous odd function of its argument. As a result of its oddity, $Q_h(\boldsymbol{x})$~identically vanishes if the set of all the non-zero weights of all the chiral multiplets in the theory consists of pairs with opposite signs; a SUSY gauge theory satisfying this
condition is called non-chiral. The EHIs studied by Rains in~\cite{Rains:2009} correspond to non-chiral gauge theories, and thus for them $Q_h=0$. For an EHI with non-zero $Q_h$ see~\cite{Ardehali:2015c}; Fig.~5 of that work shows the plot of~$Q_h$ for that example.

When all $x_i$ are small enough, so that the absolute value of all the arguments of the~$\kappa$ functions in~$Q_h$ are less than~$1$, we can use $\kappa(x)=2x^3-3x|x|+x$ to simplify~$Q_h$. The resulting expression~-- which equals~$Q_h$ for $x_i$ small enough~-- can then be considered as defining a function~$\tilde{Q}_{S^3}(x)$ for any $x_i\in\mathbb{R}$. Explicitly, we have
\begin{gather*}
\tilde{Q}_{S^3}(\boldsymbol{x})=-\frac{1}{4}\sum_j\sum_{\rho^j \in\Delta_j}\langle\rho^j\cdot \boldsymbol{x}\rangle|\langle\rho^j\cdot \boldsymbol{x}\rangle|,
\end{gather*}
with no linear term thanks to the anomaly cancellation condition (\ref{eq:anomalyCon1}), and no cubic term because of equation~(\ref{eq:sumRho=0}). In particular, $\tilde{Q}_{S^3}$ is homogeneous.

\begin{figure}[t]\centering
 \includegraphics[scale=.8]{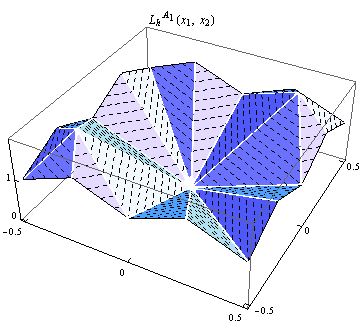}
\caption{The Rains function of SU(3) SQCD---referred to as the $A_1$ SU(3) theory in \cite{Ardehali:2015c}~-- with $N_f>3$ flavors.}\label{fig:A1}
\end{figure}

\begin{figure}[t]\centering
 \includegraphics[scale=.8]{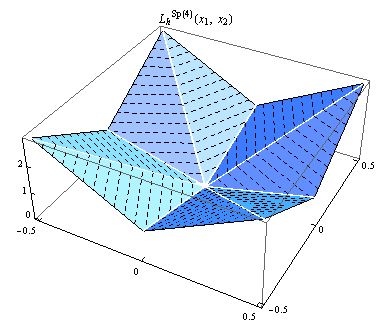}
\caption{The Rains function of Sp(4) SQCD with $N_f>3$.}\label{fig:Sp4}
\end{figure}

The real function $L_h$, which we will refer to as the \emph{Rains function} of the gauge theory, determines the effective potential $V^{\mathrm{eff}}(\boldsymbol{x};b,\beta)$. It is \emph{piecewise linear}; the quadratic terms in it cancel thanks to the anomaly cancellation condition (\ref{eq:anomalyCon3}):
\begin{gather*}
\frac{\partial^2 L_h(\boldsymbol{x})}{\partial x_l\partial x_m} =\sum_j(r_j-1)\sum_{\rho^j\in\Delta_j}\rho^j_l\rho^j_m+\sum_{\alpha\in\Delta_G}\alpha_{l}\alpha_{m}=0.
\end{gather*}
Also, $L_h$ is continuous, is even under $\boldsymbol{x}\to -\boldsymbol{x}$, and vanishes at $\boldsymbol{x}=0$; these properties follow from the properties of the function $\vartheta(x)$ defined in (\ref{eq:varthetaDef}). This function has been analyzed by Rains \cite{Rains:2009} in the context of the EHIs associated to ${\rm SU}(N)$ and ${\rm Sp}(N)$ SQCD theories. For the rank two cases considered in~\cite{Rains:2009}, this function is plotted in Figs.~\ref{fig:A1} and~\ref{fig:Sp4}.

When all $x_i$ are small enough, such that the absolute value of the argument of every $\vartheta$ function in $L_h$ is smaller than $1$, we can use $\vartheta(x)=|x|-x^2$ to simplify $L_h$. The resulting expression~-- which equals $L_h$ for small~$x_i$~-- can then be considered as defining a function $\tilde{L}_{S^3}(\boldsymbol{x})$ for any $x_i\in\mathbb{R}$. Explicitly, we have
\begin{gather}
\tilde{L}_{S^3}(\boldsymbol{x})=\frac{1}{2}\sum_j(1-r_j)\sum_{\rho^j
\in\Delta_j}|\langle\rho^j\cdot
\boldsymbol{x}\rangle|-\sum_{\alpha_+\in\Delta_G}|\langle\alpha_+\cdot
\boldsymbol{x}\rangle|.\label{eq:hypURains}
\end{gather}
Note that there is no quadratic term in $\tilde{L}_{S^3}$, thanks to the consistency condition (\ref{eq:anomalyCon3}). In particular, $\tilde{L}_{S^3}$ is homogeneous.

The content of this subsection first appeared in \cite{Ardehali:2015c}.

\subsection{The answer for non-chiral theories}

In this subsection we explain how the analysis of \cite{Ardehali:2015c} was improved in \cite{Ardehali:thesis} for the special class of non-chiral theories.
\begin{Definition}\label{definition3} A {\it non-chiral SUSY gauge theory with ${\rm U}(1)$ ${\rm R}$-symmetry} is a SUSY gauge theory with U(1) R-symmetry (as in Definition~\ref{definition1}) in which the non-zero weights in $\cup_{j}\Delta_j$ come in pairs with opposite signs. We denote the positive weights therein by $\rho_+$.
\end{Definition}
The SQCD theories discussed in Section~\ref{sec:review} are examples of non-chiral theories. We refer to the corresponding EHIs as non-chiral EHIs. All the EHIs studied by Rains in \cite{Rains:2009} are non-chiral. In non-chiral theories $\Theta=Q_h=0$, so the analysis of the hyperbolic limit simplifies.

In this subsection we use estimates more precise than the ones in the previous subsection. Therefore the symbol $\simeq$ will make an appearance.

The asymptotics of the integrand of (\ref{eq:LagEquivIndex}) can be obtained from the estimates in (\ref{eq:GammaAsyPosNegZ2}). With the aid of (\ref{eq:PochAsy}) and (\ref{eq:GammaAsyPosNegZ2}) we find the $\beta\to0$ asymptotics of $\mathcal{I}$ as
\begin{gather}
\mathcal I(b,\beta)\simeq \frac{1}{|W|} \left(\frac{2\pi}{\beta}\right)^{r_G}e^{-\mathcal{E}^{\rm DK}_0(b,\beta)}W_0(b)e^{\beta
E_{\mathrm{susy}}(b)} \int_{\mathfrak{h}_{\rm cl}} \mathrm{d}^{r_G}x\,
e^{-V^{\mathrm{eff}}(\boldsymbol{x};b,\beta)}W(\boldsymbol{x};b,\beta),\label{eq:LagIndexSimp1Th}
\end{gather}
with
\begin{gather*}
E_{\mathrm{susy}}(b)=\frac{1}{6}\left(\frac{b+b^{-1}}{2}\right)^3\operatorname{Tr}
R^3-\left(\frac{b+b^{-1}}{2}\right)\left(\frac{b^2+b^{-2}}{24}\right)\operatorname{Tr}R,
\end{gather*}
known as the {\it supersymmetric Casimir energy} (see, e.g., \cite{Brunner:2017ce}), and the expressions
\begin{gather}
\operatorname{Tr}R:=\dim G+\sum_j(r_j-1)\dim \mathcal{R}_j,\nonumber\\
\operatorname{Tr}R^3:=\dim G+\sum_j(r_j-1)^3\dim \mathcal{R}_j,\label{eq:TrR}
\end{gather}
known as the ${\rm U}(1)_R$-gravity$^2$ and ${\rm U}(1)^3_R$ 't~Hooft anomalies of the SUSY gauge theory. (The trace is over the chiral fermions in the field theory formulation: in the vector multiplet the gaugino has R-cahrge $1$, and in a chiral multiplet with R-charge $r_j$ the fermion has R-charge $r_j-1$.) We have also defined $W_0(b)$, and the \emph{real} function $W(\boldsymbol{x};b,\beta)$ via
\begin{gather}
W_0(b)=\prod_j\prod_{\rho^j=0}\Gamma_h(r_j\omega),\label{eq:W0Def}\\
W(\boldsymbol{x};b,\beta)=\prod_j
\prod_{\rho^j_+}\frac{\psi_b\big({-}\frac{2\pi
i}{\beta}\{\langle\rho^j_+\cdot\boldsymbol{x}\rangle\}+(r_j-1)\frac{b+b^{-1}}{2}\big)}{\psi_b\big({-}\frac{2\pi
i}{\beta}\{\langle\rho^j_+\cdot\boldsymbol{x}\rangle\}-(r_j-1)\frac{b+b^{-1}}{2}\big)}\nonumber\\
\hphantom{W(\boldsymbol{x};b,\beta)=}{} \times
\prod_{\alpha_+}\frac{\psi_b\big({-}\frac{2\pi
i}{\beta}\{\langle\alpha_+\cdot\boldsymbol{x}\rangle\}+\frac{b+b^{-1}}{2}\big)}{\psi_b\big({-}\frac{2\pi
i}{\beta}\{\langle\alpha_+\cdot\boldsymbol{x}\rangle\}-\frac{b+b^{-1}}{2}\big)}.\label{eq:W+Def}
\end{gather}
In (\ref{eq:W0Def}), the second product is over the zero weights of $\mathcal{R}_j$ (the adjoint representation, for example, has $r_G$ such weights), and $\omega$ is defined as $\omega:=i\big(b+b^{-1}\big)/2$. The~$\rho^j_+$ in~(\ref{eq:W+Def}) denote the positive weights of~$\mathcal{R}_j$.

That $W(\boldsymbol{x};b,\beta)$ is real follows from~(\ref{eq:hyperbolicGammaConj}) and~(\ref{eq:hyperbolicGammaPsi}).

Our claim in (\ref{eq:LagIndexSimp1Th}) that \emph{the matrix-integral is approximated well with the integral of its approximate integrand} is justified because the estimates we have used inside the integrand are uniform and accurate up to exponentially small corrections of the type~$e^{-1/\beta}$.

Now, from (\ref{eq:hyperbolicGammaConj}) it follows that $W_0(b)$ is a real number; it is moreover nonzero and finite, because with the assumption $r_j\in{}]0,2[$ the arguments $r_j \omega$ avoid the zeros and poles of the hyperbolic gamma function as described in Section~\ref{sec:defs}. We would thus make an $O\big(\beta^0\big)$ error in the asymptotics of $\log\mathcal{I}(b,\beta)$ if in~(\ref{eq:LagIndexSimp1Th}) we set $W_0(b)$, along with~$|W|$ and $e^{\beta E_{\mathrm{susy}}(b)}$, to unity. In other words,
\begin{gather}
\mathcal I(b,\beta)\approx \left(\frac{2\pi}{\beta}\right)^{r_G}e^{-\mathcal{E}^{\rm DK}_0(b,\beta)}
\int_{\mathfrak{h}_{\rm cl}} \mathrm{d}^{r_G}x\, e^{-V^{\mathrm{eff}}(\boldsymbol{x};b,\beta)}W(\boldsymbol{x};b,\beta),\label{eq:LagIndexSimpO1}
\end{gather}
with an $O\big(\beta^0\big)$ error upon taking the logarithms of the two sides.

We are hence left with the asymptotic analysis of the integral $\int_{\mathfrak{h}_{\rm cl}} e^{-V}W$. From here, standard methods of asymptotic analysis can be employed.

Writing $V^{\mathrm{eff}}$ in terms of the Rains function $L_h$, (\ref{eq:LagIndexSimpO1}) simplifies to
\begin{gather}
\mathcal{I}(b,\beta)\approx \left(\frac{2\pi}{\beta}\right)^{r_G} e^{-\mathcal{E}^{\rm DK}_0(b,\beta)}\int_{\mathfrak{h}_{\rm cl}}
\mathrm{d}^{r_G}x\, e^{-\frac{4\pi^2}{\beta}\big(\frac{b+b^{-1}}{2}\big)L_h(\boldsymbol{x})}W(\boldsymbol{x};b,\beta).\label{eq:LagIndexSimp1noTheta}
\end{gather}

It will be useful for us to know that $W(\boldsymbol{x};b,\beta)$ is a positive semi-definite function of $\boldsymbol{x}$; this follows from (\ref{eq:hyperbolicGammaConj}) and (\ref{eq:hyperbolicGammaPsi}).

The rest of the analysis, leading to our main result in (\ref{eq:LagIndexSimp6noTheta}), involves: $i)$ showing that the integral in~(\ref{eq:LagIndexSimp1noTheta}) localizes around the locus $\mathfrak{h}_{qu}$ of minima of $L_h$, so that the leading~$O(1/\beta)$ terms in~(\ref{eq:LagIndexSimp6noTheta}) are justified, and $ii)$ rescaling, in the remaining localized integral, the distance~$\Delta \boldsymbol{x}_{\perp}$ from $\mathfrak{h}_{qu}$ as in $\Delta \boldsymbol{x}_{\perp}\to\big(\frac{\beta}{2\pi}\big)\Delta \boldsymbol{x}_{\perp}$, so that $r_G-\dim \mathfrak{h}_{qu}$ of the $(2\pi/\beta)$ pre-factors of~(\ref{eq:LagIndexSimp1noTheta}) are absorbed through integration, justifying the subleading $O\big(\log\big(\frac{2\pi}{\beta}\big)\big)$ term in~(\ref{eq:LagIndexSimp6noTheta}). The reader not interested in the details of the derivation is invited to skip the following analysis, and continue reading from~(\ref{eq:LagIndexSimp6noTheta}).

To analyze the integral in (\ref{eq:LagIndexSimp1noTheta}), first note that the integrand is not smooth over $\mathfrak{h}_{\rm cl}$. We hence break $\mathfrak{h}_{\rm cl}$ into sets on which $L_h$ is linear. These sets can be obtained as follows. Define
\begin{gather*}
\mathcal{S}_g:=\bigcup_{\alpha_+}\{\boldsymbol{x}\in \mathfrak{h}_{\rm cl}|\langle\alpha_+\cdot
\boldsymbol{x}\rangle\in \mathbb{Z}\},\qquad\mathcal{S}_j:=\bigcup_{\rho^j_+}\{\boldsymbol{x}\in\mathfrak{h}_{\rm cl}|\langle\rho^{j}_+\cdot
\boldsymbol{x}\rangle\in\mathbb{Z}\},\\
 \mathcal{S}:=\bigcup_{j}\mathcal{S}_j\cup \mathcal{S}_g.
\end{gather*}
It should be clear that everywhere in $\mathfrak{h}_{\rm cl}$, except on $\mathcal{S}$, the function $L_h$ is guaranteed to be linear~-- and therefore smooth.

The set $\mathcal{S}$ consists of a union of codimension one affine hyperplanes inside the space of the~$x_i$. These hyperplanes chop $\mathfrak{h}_{\rm cl}$ into (finitely many, convex) polytopes~$\mathcal{P}_n$. The integral in~(\ref{eq:LagIndexSimp1noTheta}) then decomposes to
\begin{gather}
\mathcal{I}(b,\beta)\approx e^{-\mathcal{E}^{\rm DK}_0(b,\beta)}\sum_n \left(\frac{2\pi}{\beta}\right)^{r_G} \int_{\mathcal{P}_n} \mathrm{d}^{r_G}x\,
e^{-\frac{4\pi^2}{\beta}\big(\frac{b+b^{-1}}{2}\big)L_h(\boldsymbol{x})}W(\boldsymbol{x};b,\beta).\label{eq:LagIndexSimp2noTheta}
\end{gather}

Let $\mathcal{S}^{(\beta)}_g$ denote the set of all points in $\mathfrak{h}_{\rm cl}$ that are at a distance less than $N_0\beta$ from $\mathcal{S}_g$, with some fixed $N_0>0$. We divide $\mathcal{P}_n$ into $i)$ $\mathcal{P}_n\cap\mathcal{S}^{(\beta)}_g$, and $ii)$ the rest of $\mathcal{P}_n$, which we denote by $\mathcal{P}'_n$. Now, by taking $N_0$ to be large enough, we can push $\mathcal{P}'_n$ away from the zeros of $\psi_b$, and thus make $w_i<W(\boldsymbol{x};b,\beta)<w_s$ over $\mathcal{P}'_n$ (with some $0<w_i$ and some $w_s<\infty$). Therefore the contribution that the $n$th summand in~(\ref{eq:LagIndexSimp2noTheta}) receives from $\mathcal{P}'_n$ is well approximated (with an~$O\big(\beta^0\big)$ error upon taking the logs) by
\begin{gather}
J_n:=\left(\frac{2\pi}{\beta}\right)^{r_G}\int_{\mathcal{P}'_n} \mathrm{d}^{r_G}x\,
e^{-\frac{4\pi^2}{\beta}\big(\frac{b+b^{-1}}{2}\big)L_h(\boldsymbol{x})}.\label{eq:ithIntegralSimplPp}
\end{gather}

Let's further replace $\mathcal{P}'_n$ in (\ref{eq:ithIntegralSimplPp}) with $\mathcal{P}_n$; we will shortly see that this replacement introduces a negligible error. We would hence like to estimate
\begin{gather}
I_n:=\left(\frac{2\pi}{\beta}\right)^{r_G}\int_{\mathcal{P}_n} \mathrm{d}^{r_G}x\,
e^{-\frac{4\pi^2}{\beta}\big(\frac{b+b^{-1}}{2}\big)L_h(\boldsymbol{x})}.\label{eq:ithIntegralSimplP}
\end{gather}

Since $L_h$ is linear on each $\mathcal{P}_n$, its minimum over $\mathcal{P}_n$ is guaranteed to be realized on~$\partial\mathcal{P}_n$. Let us assume that this minimum occurs on the $k$th $j$-face of $\mathcal{P}_n$, which we denote by $j_n$-$\mathcal{F}^k_n$. We denote the value of~$L_h$ on this $j$-face by $L_{h,\min}^n$. Equipped with this notation, we can write~(\ref{eq:ithIntegralSimplP}) as
\begin{gather}
I_n= \left(\frac{2\pi}{\beta}\right)^{r_G} e^{-\frac{4\pi^2}{\beta}\big(\frac{b+b^{-1}}{2}\big)L_{h,\min}^n}\int_{\mathcal{P}_n}
\mathrm{d}^{r_G}x\, e^{-\frac{4\pi^2}{\beta}\big(\frac{b+b^{-1}}{2}\big)\Delta L^n_h(\boldsymbol{x})},\label{eq:LagIndexSimp3noTheta}
\end{gather}
where $\Delta L^n_h(\boldsymbol{x}):=L_h(\boldsymbol{x})-L_{h,\min}^n$ is a linear function on~$\mathcal{P}_n$. Note that $\Delta L^n_h(\boldsymbol{x})$ vanishes on~\smash{$j_n$-$\mathcal{F}^k_n$}, and it increases as we go away from $j_n$-$\mathcal{F}^k_n$ and into the interior of $\mathcal{P}_n$. (The last sentence, as well as the rest of the discussion leading to~(\ref{eq:LagIndexSimp6noTheta}), would receive a trivial modification if $j_n=r_G$ (corresponding to constant $L_h$ over $\mathcal{P}_n$).) Therefore as $\beta\to 0$, the integral in~(\ref{eq:LagIndexSimp3noTheta}) localizes around $j_n$-$\mathcal{F}^k_n$.

To further simplify (\ref{eq:LagIndexSimp3noTheta}), we now adopt a~set of new coordinates~-- affinely related to $x_i$ and with unit Jacobian~-- that are convenient on $\mathcal{P}_n$. We pick a point on $j_n$-$\mathcal{F}^k_n$ as the new origin, and parameterize $j_n$-$\mathcal{F}^k_n$ with $\bar{x}_1,\dots,\bar{x}_{j_n}$. We take $x_{\mathrm{in}}$ to parameterize a direction perpendicular to all the $\bar{x}$s, and to increase as we go away from $j_n$-$\mathcal{F}^k_n$ and into the interior of $\mathcal{P}_n$. Finally, we pick $\tilde{x}_1,\dots,\tilde{x}_{r_G-j_n-1}$ to
parameterize the perpendicular directions to $x_{\mathrm{in}}$ and the $\bar{x}$s. Note that, because $\Delta L^n_h$ is linear on~$\mathcal{P}_n$, it does not depend on the $\bar{x}$s; they parameterize its flat directions. By re-scaling $\bar{x}$, $x_{\mathrm{in}}$, $\tilde{x}\mapsto
\frac{\beta}{2\pi}\bar{x}$, $\frac{\beta}{2\pi}x_{\mathrm{in}}$, $\frac{\beta}{2\pi}\tilde{x}$, we can absorb the $\big(\frac{2\pi}{\beta}\big)^{r_G}$ factor in~(\ref{eq:LagIndexSimp3noTheta}) into the integral, and write the result as
\begin{gather}
I_n=\int_{\frac{2\pi}{\beta}\mathcal{P}_n} \mathrm{d}^{j_n}\bar{x}\,
\mathrm{d}x_{\mathrm{in}}\, \mathrm{d}^{r_G-j_n-1}\tilde{x}\, e^{-2\pi\big(\frac{b+b^{-1}}{2}\big)\Delta
L^n_h(x_{\mathrm{in}},\mathbf{\tilde{x}})}.\label{eq:ithIntegralSimpl1}
\end{gather}
To eliminate $\beta$ from the exponent, we have used the fact that $\Delta L^n_h$ depends homogenously on the new coordinates. We are also denoting the re-scaled polytope schematically by $\frac{2\pi}{\beta}\mathcal{P}_n$.

Instead of integrating over all of $\frac{2\pi}{\beta}\mathcal{P}_n$ though, we can restrict to $x_{\mathrm{in}}<\epsilon/\beta$ with some small $\epsilon>0$. The reason is that the integrand of~(\ref{eq:ithIntegralSimpl1}) is exponentially suppressed (as~$\beta\to 0$) for $x_{\mathrm{in}}>\epsilon/\beta$. We take $\epsilon>0$ to be small enough such that a hyperplane at $x_{\mathrm{in}}=\epsilon/\beta$, and parallel to $j_n$-$\mathcal{F}^k_n$, cuts off a prismatoid $P^n_{\epsilon/\beta}$ from $\frac{2\pi}{\beta}\mathcal{P}_n$. After restricting the integral in~(\ref{eq:ithIntegralSimpl1}) to $P^n_{\epsilon/\beta}$, the integration over the $\bar{x}$s is easy to perform. The only potential difficulty is that the range of the $\bar{x}$ coordinates may depend on $x_{\mathrm{in}}$ and the~$\tilde{x}s$. But since we are dealing with a prismatoid, the dependence is linear, and by the time the range is modified significantly (compared to its $O(1/\beta)$ size on the re-scaled $j$-face $\frac{2\pi}{\beta}(j_n$-$\mathcal{F}^k_n)$), the integrand is exponentially suppressed. Therefore we can neglect the dependence of the range of the $\bar{x}$s on the other coordinates in~(\ref{eq:ithIntegralSimpl1}). The integral then simplifies to
\begin{gather}
I_n\approx\left(\frac{2\pi}{\beta}\right)^{j_n}\, \operatorname{vol}\big(j_n\text{-}\mathcal{F}^k_n\big)\int_{\hat{P}^n_{\epsilon/\beta}}
\mathrm{d}x_{\mathrm{in}}\, \mathrm{d}^{r_G-j_n-1}\tilde{x}\, e^{-2\pi\big(\frac{b+b^{-1}}{2}\big)\Delta L^n_h(x_{\mathrm{in}},\mathbf{\tilde{x}})},\label{eq:ithIntegralSimpl2}
\end{gather}
where $\hat{P}^n_{\epsilon/\beta}$ is the pyramid obtained by restricting $P^n_{\epsilon/\beta}$ to $\bar{x}_1=\dots =\bar{x}_{j_n}=0$. The logarithms of the two sides of~(\ref{eq:ithIntegralSimpl2}) differ by $O(\beta)$, with the error mainly arising from our neglect of the possible dependence of the range of the $\bar{x}$ coordinates in~(\ref{eq:ithIntegralSimpl1}) on $x_{\mathrm{in}}$ and the $\tilde{x}s$. (Recall that the other error, arising from restricting the integral in~(\ref{eq:ithIntegralSimpl1}) to $P^n_{\epsilon/\beta}$, is exponentially small.)

We now take $\epsilon\to\infty$ in (\ref{eq:ithIntegralSimpl2}). This introduces an exponentially small error, as the integrand is exponentially suppressed (as $\beta\to 0$) for $x_{\mathrm{in}}>\epsilon/\beta$. The resulting integral is strictly positive, because it is the integral of a~strictly positive function. We denote by $A_n$ the result of the integral multiplied by $\operatorname{vol}\big(j_n\text{-}\mathcal{F}^k_n\big)$. Then~$I_n$ can be approximated as
\begin{gather}
I_n\approx e^{-\frac{4\pi^2}{\beta}\big(\frac{b+b^{-1}}{2}\big)L_{h,\min}^n}\left(\frac{2\pi}{\beta}\right)^{j_n} A_n. \label{eq:LagIndexSimp4noTheta}
\end{gather}

We are now in a position to argue $J_n\approx I_n$. If we had integrated over $\mathcal{P}'_n$, then we would end up with an expression similar to (\ref{eq:LagIndexSimp4noTheta}), in which $L_{h,\min}^n$ would be replaced with the minimum of~$L_h$ over~$\mathcal{P}'_n$; but since~$L_h$ is piecewise linear, the difference between the new minimum and~$L_{h,\min}^n$ would be $O(\beta)$, which translates to an~$O\big(\beta^0\big)$ multiplicative difference between~$J_n$ and~$I_n$. Other sources of difference between $J_n$ and $I_n$ similarly introduce negligible error; more
precisely, we have $\log I_n=\log J_n+O\big(\beta^0\big)$.

The dominant contribution to $\mathcal{I}(b{,}\beta)$ comes, of course, from the terms/polytopes whose~$L_{h{,}\min}^n$ is smallest. If these polytopes are labeled by $n=n_\ast^1,n_\ast^2,\dots$, we can introduce $\mathfrak{h}_{qu}$ and $\dim \mathfrak{h}_{qu}$ via
\begin{gather}
\mathfrak{h}_{qu}:=\bigcup_{n_\ast} j_{n_\ast}\text{-}\mathcal{F}^k_{n_\ast},\qquad
\dim \mathfrak{h}_{qu}:=\max (j_{n_\ast}),\label{eq:dimhquDef}
\end{gather}
with $j_{n_\ast}$ the dimensions of the $j$-faces with minimal $L_{h,\min}^n$.

Put colloquially, if $\mathfrak{h}_{qu}$ has multiple connected components, by $\dim \mathfrak{h}_{qu}$ we mean the dimension of the component(s) with greatest dimension, while if a connected component consists of several intersecting flat elements inside $\mathfrak{h}_{\rm cl}$, by its dimension we mean the dimension of the flat element(s) of maximal dimension.

Our final estimate for the contribution to $\mathcal{I}(b,\beta)$ from $\cup_n\mathcal{P}'_n$ is thus
\begin{gather}
Be^{-\mathcal{E}^{\rm DK}_0(b,\beta)-\frac{4\pi^2}{\beta}\big(\frac{b+b^{-1}}{2}\big)L_{h,\min}}\left(\frac{2\pi}{\beta}\right)^{\dim \mathfrak{h}_{qu}}, \label{eq:LagIndexSimp5noTheta}
\end{gather}
where $L_{h,\min}:=L_{h,\min}^{n_\ast}$, and $B$ is some positive real number.

We are left with determining the contribution to $\mathcal{I}(b,\beta)$ coming from $\mathcal{S}^{(\beta)}_g$. Over \smash{$\mathcal{P}_n\cap\mathcal{S}^{(\beta)}_g$}, the simple estimate $W(\boldsymbol{x};b,\beta)= O(1)$ (which follows from the fact that $W(\boldsymbol{x};b,\beta)$ is uniformly bounded on $\mathcal{S}^{(\beta)}_g$) suffices for our purposes; we thus learn that the contribution that the integral~(\ref{eq:LagIndexSimp2noTheta}) receives from $\mathcal{P}_n\cap\mathcal{S}^{(\beta)}_g$ is not only positive, but also
\begin{gather*}
O\left(\int_{\frac{2\pi}{\beta}\big(\mathcal{P}_n\cap\mathcal{S}^{(\beta)}_g\big)}
\mathrm{d}^{j_n}\bar{x}\, \mathrm{d}x_{\mathrm{in}}\, \mathrm{d}^{r_G-j_n-1}\tilde{x}\, e^{-2\pi\big(\frac{b+b^{-1}}{2}\big)\Delta
L^n_h(x_{\mathrm{in}},\mathbf{\tilde{x}})}\right).
\end{gather*}

Now, the argument of the $O$ above is nothing but the difference between $I_n$ and $J_n$, which we already argued to be negligible. Thus the contribution to $\mathcal{I}(b,\beta)$ coming from $\mathcal{S}^{(\beta)}_g$ is negligible.

Using the explicit expression (\ref{eq:EdkEquiv}) for $\mathcal{E}^{\rm DK}_0(b,\beta)$ (which can be rewritten in terms of $\operatorname{Tr}R$ in equation~(\ref{eq:TrR})), and noting that (\ref{eq:LagIndexSimp5noTheta}) is an accurate estimate for $\mathcal{I}(b,\beta)$ up to a multiplicative factor of order $\beta^0$ (but not order $\beta$), we arrive at the following proposition as our main result.
\begin{Proposition}\label{proposition} The Romelsberger index \eqref{eq:LagEquivIndex} of a non-chiral SUSY gauge theory with ${\rm U}(1)$ ${\rm R}$-symmetry $($as in Definition~{\rm \ref{definition3})} has the following asymptotics in the hyperbolic limit:
\begin{gather}
\boxed{\log \mathcal{I}(b,\beta)= -\frac{\pi^2}{3\beta}\left(\frac{b+b^{-1}}{2}\right)(\operatorname{Tr}R+12L_{h,\min})
+\dim \mathfrak{h}_{qu}\log\left(\frac{2\pi}{\beta}\right)+O\big(\beta^0\big),\label{eq:LagIndexSimp6noTheta}
}
\end{gather}
with $\dim \mathfrak{h}_{qu}$ defined precisely in~\eqref{eq:dimhquDef}.
\end{Proposition}

\section[The minimization problem via generalized triangle inequalities]{The minimization problem\\ via generalized triangle inequalities}\label{sec:gti}

To the order shown in (\ref{eq:LagIndexSimp6noTheta}), the hyperbolic asymptotics of $\mathcal{I}(b,\beta)$ is determined by the minimum value and the dimension of the locus of minima of the Rains function. Finding these two numbers involves solving a~minimization problem for~$L_h$. This can be done only on a~case-by-case basis at the moment. The tool allowing us to solve the minimization problem is often Lemma~3.2 of~\cite{Rains:2009}, stating that for any sequence of real numbers $c_1,\dots, c_n$, $d_1,\dots, d_n$, the following inequality holds:
\begin{gather}
\sum_{1\le i,j\le n}\vartheta(c_i-d_j)-\sum_{1\le i<j\le n}\vartheta(c_i-c_j)-\sum_{1\le i<j\le n}\vartheta(d_i-d_j)\ge
\vartheta\left(\sum_{1\le i\le n}(c_i-d_i)\right),\label{eq:RainsGTI}
\end{gather}
with equality iff the sequence can be permuted so that either
\begin{gather*}
\{c_1\}\le\{d_1\}\le\{c_2\}\le\cdots\le \{d_{n-1}\}\le\{c_n\}\le\{d_n\},
\end{gather*}
or
\begin{gather*}
\{d_1\}\le\{c_1\}\le\{d_2\}\le\cdots\le \{c_{n-1}\}\le\{d_n\}\le\{c_n\}.
\end{gather*}
The proof can be found in \cite{Rains:2009}.

Re-scaling with $c_i,d_i\mapsto vc_i,vd_i$, taking $v\to 0^+$, and using the relation $\vartheta(vx)=v|x|-v^2 x^2$ (which holds for small enough $v$), Rains obtains the following corollary of~(\ref{eq:RainsGTI}):
\begin{gather}
\sum_{1\le i,j\le n}|c_i-d_j|-\sum_{1\le i<j\le n}|c_i-c_j|-\sum_{1\le i<j\le n}|d_i-d_j|\ge \left|\sum_{1\le i\le n}(c_i-d_i)\right|,\label{eq:RainsGTIav}
\end{gather}
with equality iff the sequence can be permuted so that either
\begin{gather*}
c_1\le d_1 \le c_2 \le\cdots\le d_{n-1} \le c_n \le d_n,
\end{gather*}
or
\begin{gather*}
d_1 \le c_1 \le d_2 \le\cdots\le c_{n-1} \le d_n \le c_n.
\end{gather*}

The fact that the inequality (\ref{eq:RainsGTIav}) arise as a~corollary of~(\ref{eq:RainsGTI}) justifies the name ``generalized triangle inequality'' for the latter.

We now consider four examples of non-chiral EHIs for which the minimization problem of $L_h$ can be fully addressed.

\subsection[${\rm SU}(N_c)$ SQCD with $N_f>N_c$ flavors]{$\boldsymbol{{\rm SU}(N_c)}$ SQCD with $\boldsymbol{N_f>N_c}$ flavors}\label{subsec:sqcd}

This theory was described in Section~\ref{sec:review}. Its corresponding EHI is
\begin{gather}
\mathcal{I}_{N_c,N_f}(b,\beta)=\frac{(p;p)^{N_c-1}(q;q)^{N_c-1}}{N_c!}\int \mathrm{d}^{N_c-1}x \frac{\prod\limits_{i=1}^{N_c}\Gamma^{N_f}\big((pq)^{r_f/2} z_i^{\pm 1}\big)}{\prod\limits_{1\le i<j\le N_c} \Gamma\big((z_i/z_j)^{\pm 1}\big)},\label{eq:SQCDindex}
\end{gather}
with $r_f=1-\frac{N_c}{N_f}$, and $\prod\limits_{i=1}^{N_c} z_i=1$. Recall that this is essentially the same EHI as $I^{(m)}_{A_n}$ of~\cite{Rains:2005} with $n=N_c-1$ and $m=N_f-N_c-1$.

The Rains function of the theory is
\begin{gather}
L_h^{N_c,N_f}(x_1,\dots,x_{N_c-1}) =N_f (1-r_f)\sum_{i=1}^{N_c}\vartheta(x_i)-\sum_{1\le i<j\le N_c}\vartheta(x_i-x_j)\nonumber\\
\hphantom{L_h^{N_c,N_f}(x_1,\dots,x_{N_c-1})}{} =N_c\sum_{i}\vartheta(x_i)-\sum_{1\le i<j\le N_c}\vartheta(x_i-x_j).\label{eq:AkRains}
\end{gather}
The $x_{N_c}$ in the above expression is constrained by $\sum\limits_{i=1}^{N_c} x_i\in\mathbb{Z}$, although since $\vartheta(x)$ is periodic with period one we can simply replace $x_{N_c}\to -x_1-\dots-x_{N_c-1}$. For $N_c=3$ the resulting function is illustrated in Fig.~\ref{fig:A1}.

We recommend that the reader convince herself that the Rains function in (\ref{eq:AkRains}) can be easily written down by examining the integrand of (\ref{eq:SQCDindex}). Whenever the Romelsberger index (or the EHI) of a SUSY gauge theory is available in the literature, a similar examination of the integrand quickly yields the theory's $L_h$ and $Q_h$ functions.

Using Rains' generalized triangle inequality (\ref{eq:RainsGTI}), in the special case where $d_i=0$, we find that the above function is minimized when all $x_i$ are zero. Therefore
\begin{gather*}
L^{N_c,N_f}_{h,\min}=0,\qquad \dim \mathfrak{h}^{N_c,N_f}_{qu}=0.
\end{gather*}

A similar story applies to the ${\rm Sp}(2N)$ SQCD theory discussed in Section~\ref{sec:review}. We invite the interested reader to reproduce the plot of the Rains function of the ${\rm Sp}(4)$ SQCD for $N_f>3$ shown in Fig.~\ref{fig:Sp4}. (The Romelsberger index of the ${\rm Sp}(2N)$ SQCD theories can be found in~\cite{Dolan:2008}. Lemma~3.3 of Rains \cite{Rains:2009} establishes that $L_h^{Sp(2N)}(\boldsymbol{x})$ is minimized only at $\boldsymbol{x}=0$, for any $N_f>N+1$.)

\subsubsection*{All-orders asymptotics}

A more careful study shows \cite{Ardehali:2015c,Rains:2009}
\begin{gather*}
\log \mathcal{I}_{N_c,N_f}(b,\beta)\sim -\frac{\pi^2}{3\beta}\left(\frac{b+b^{-1}}{2}\right)\operatorname{Tr}R +
\log Z^{N_c,N_f}_{S^3}(b)+\beta E_{\mathrm{susy}}(b),\qquad \text{as} \ \beta\to 0\label{eq:SQCDindexAsy3}
\end{gather*}
with
\begin{gather*}
Z^{N_c,N_f}_{S^3}(b)=\frac{1}{N_c!} \int \mathrm{d}^{N_c-1}x \frac{\prod\limits_{i=1}^{N_c}\Gamma_h^{N_f}(r_f \omega\pm
x_i)}{\prod\limits_{1\le i<j\le N_c}\Gamma_h(\pm (x_i-x_j))},\label{eq:SQCD3dZ}
\end{gather*}
where $\omega:=(\omega_1+\omega_2)/2$, and the integral is over $x_1,\dots,x_{N_c-1}\in{}]{-}\infty,\infty[$.

The content of this subsection is essentially due to Rains~\cite{Rains:2009}.

\subsection[${\rm SO}(2N+1)$ SQCD with $N_f>2N-1$]{$\boldsymbol{{\rm SO}(2N+1)}$ SQCD with $\boldsymbol{N_f>2N-1}$}\label{subsec:soN}

For $G=$${\rm SO}(n)$, and $n_\chi=N_f$ chiral multiplets of R-charge $r=1-\frac{n-2}{N_f}$ in the vector representation of ${\rm SO}(n)$, we
get the ${\rm SO}(n)$ SQCD theory with $N_f$ flavors. For the R-charges to be greater than zero, and the gauge group to be semi-simple, we require $0<n-2<N_f$.

We consider odd $n$. The analysis for even $n$ is similar. The EHI of the ${\rm SO}(2N+1)$ SQCD with $N_f$ flavors reads
\begin{gather}
\mathcal{I}_{{\rm SO}(2N+1)}(b,\beta) =\frac{(p;p)^N(q;q)^N}{2^N N!}\Gamma^{N_f}\big((pq)^{r/2}\big)\nonumber\\
\hphantom{\mathcal{I}_{{\rm SO}(2N+1)}(b,\beta) =}{} \times\int \mathrm{d}^N x\,
\frac{\prod\limits_{j=1}^{N}\Gamma^{N_f}\big((pq)^{r/2}z_j^{\pm1}\big)}{\prod\limits_{j=1}^{N}
\Gamma\big(z_j^{\pm1}\big)\prod\limits_{i<j}\big(\Gamma\big((z_i z_j)^{\pm1}\big)\Gamma\big((z_i/
z_j)^{\pm1}\big)\big)}.\label{eq:SONindex}
\end{gather}

\begin{figure}[t]\centering
 \includegraphics[scale=.7]{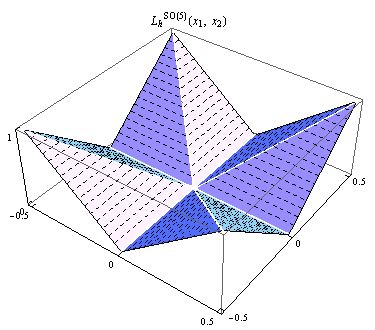}
\caption{The Rains function of the ${\rm SO}(5)$ SQCD theory with $N_f>3$.}\label{fig:SO5}
\end{figure}

The Rains function of the theory is
\begin{gather}
L_h^{{\rm SO}(2N+1)}(\boldsymbol{x})=(2N-2)\sum_{j=1}^{N}\vartheta(x_j)
-\sum_{1\le i<j\le N}\vartheta(x_i+x_j)-\sum_{1\le i<j\le N}\vartheta(x_i-x_j).\label{eq:soNRains}
\end{gather}
For the case $N=2$, corresponding to the ${\rm SO}(5)$ theory, this function is illustrated in Fig.~\ref{fig:SO5}.

To find the minima of the above function, we need the following result. For $-1/2\le x_i\le 1/2$,
\begin{gather}
(2N-2)\sum_{1\le j\le N}\vartheta(x_j)-\sum_{1\le i<j\le
N}\vartheta(x_i+x_j)-\sum_{1\le i<j\le N}\vartheta(x_i-x_j) \nonumber\\
\qquad{} =2\sum_{1\le i<j\le N}\min (|x_i|,|x_j|)\nonumber\\
\qquad{} =2(N-1)\min (|x_i|)+2(N-2)\min _2(|x_i|)+\cdots +2\min _{N-1}(|x_i|),\label{eq:SONvarthetaId}
\end{gather}
where $\min (|x_i|)$ stands for the smallest of $|x_1|,\dots,|x_N|$, while $\min _2(|x_i|)$ stands for the next to smallest element, and so on. To prove~(\ref{eq:SONvarthetaId}), one can first verify it for $N=2$, and then use induction for $N>2$.

Applying (\ref{eq:SONvarthetaId}) we find that the Rains function in (\ref{eq:soNRains}) is minimized to zero when one (and only one) of the $x_j$ is nonzero, and the rest are zero. This follows from the fact that $\max (|x_i|)$ does not show up on the r.h.s.\ of~(\ref{eq:SONvarthetaId}). Therefore we have
\begin{gather*}
L^{{\rm SO}(2N+1)}_{h,\min}=0,\qquad \dim \mathfrak{h}^{{\rm SO}(2N+1)}_{qu}=1.
\end{gather*}

\subsubsection*{More precise asymptotics}

A careful study shows \cite{Ardehali:2015c}
\begin{gather}
\log \mathcal{I}_{{\rm SO}(2N+1)}(b,\beta)= -\mathcal{E}_0^{\rm DK}(b,\beta)+\log\left(\frac{2\pi}{\beta}\right)+\log
Y_{3d}(b)+o(1)\qquad \text{as} \ \beta\to 0\label{eq:SONindexAsy2}
\end{gather}
where
\begin{gather}
Y_{3d}(b)=\frac{\Gamma_h^{N_f}(\omega r)}{2^N (N-1)!} \nonumber\\
\hphantom{Y_{3d}(b)=}{} \times \int \mathrm{d}^{N-1} x\, \frac{\prod\limits_{j=1}^{N-1}\Gamma_h^{N_f}(\omega r
\pm x_j)}{\prod\limits_{j=1}^{N-1} \Gamma_h(\pm x_j)\prod\limits_{1\le i<j\le
N-1}(\Gamma_h(\pm (x_i+x_j))\Gamma_h(\pm (x_i-x_j)))},\label{eq:Yso3}
\end{gather}
with the integral over $x_1,\dots,x_{N-1}\in{}]{-}\infty,\infty[$.

The hyperbolic reduction (\ref{eq:Yso3}) of the EHI (\ref{eq:SONindex}) is unusual, compared with the cases studied by Rains \cite{Rains:2009} for which the hyperbolic reduction has essentially the same integrand as the elliptic integral but with hyperbolic gamma functions replacing elliptic gamma functions.

\subsubsection*{All-orders asymptotics for the $\boldsymbol{{\rm SO}(3)}$ theory with $\boldsymbol{N_f=2}$ when $\boldsymbol{b=1}$}

Interestingly, for this case an indirect approach (through the machinery of \cite{Bourdier:2015}) can be used to complete the asymptotic expansion in (\ref{eq:SONindexAsy2}) to all orders, with the result reading~\cite{Ardehali:2015c}
\begin{gather*}
\log \mathcal{I}_{{\rm SO}(3)}(\beta)\sim \log\left(\frac{\pi}{2\beta}-\frac{1}{2\pi}\right)+\frac{3}{8}\beta\qquad
\text{as} \ \beta\to 0.\label{eq:SO3indexAsy3}
\end{gather*}
Note that for the ${\rm SO}(3)$ theory with $N_f=2$ we have $\operatorname{Tr} R=0$, so that the generically-leading~$O(1/\beta)$ term (i.e.,
$-\mathcal{E}_0^{\rm DK}(b,\beta)$) vanishes.

The content of this subsection first appeared in~\cite{Ardehali:2015c}.

\subsection[Puncture-less ${\rm SU}(2)$ class-$\mathcal{S}$ quivers]{Puncture-less $\boldsymbol{{\rm SU}(2)}$ class-$\boldsymbol{\mathcal{S}}$ quivers}\label{subsec:classS}

An interesting class of SUSY gauge theories with U(1) R-symmetry arises from quiver gauge theories associated to Riemann surfaces of genus $g\ge2$, without punctures. They are referred to as class-$\mathcal{S}$ theories, and they bridge EHIs to topological QFT on Riemann surfaces~\cite{Gadde:2009tqft}. (See also~\cite{Teschner:2014} for a discussion of these theories (as~$A_1$ theories of class~$\mathcal{S}$) in the context of the celebrated AGT correspondence, and~\cite{Beem:2012ca} for a discussion of their Romelsberger index.)

\begin{figure}[t]\centering
 \includegraphics[scale=1.5]{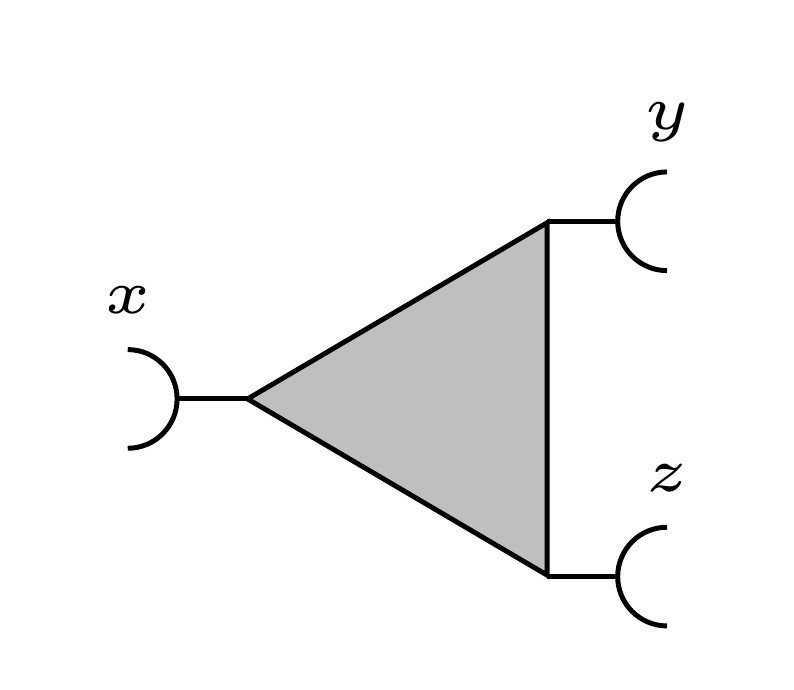}
\caption{The building block of the puncture-less SU(2) class-$\mathcal{S}$ theories.} \label{fig:classSb}
\end{figure}

\begin{figure}[t]\centering
 \includegraphics[scale=1]{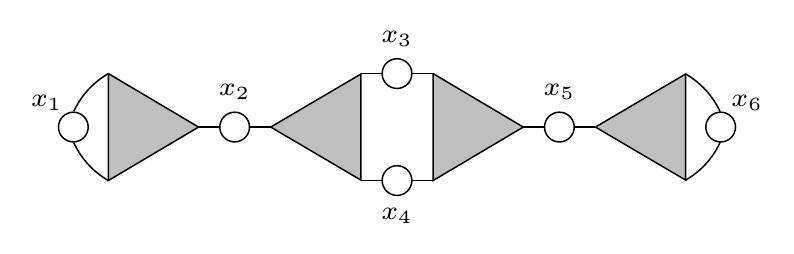}
\caption{The quiver diagram of the $g=3$ class-$\mathcal{S}$ theory.}\label{fig:g3S}
\end{figure}

These quivers can be constructed from fundamental blocks of the kind shown in Fig.~\ref{fig:classSb}. The triangle in Fig.~\ref{fig:classSb} represents eight chiral multiplets of R-charge $2/3$ transforming in the tri-fundamental representation of the three ${\rm SU}(2)$ gauge groups represented by the (semi-circular) nodes; more precisely, when two semi-circular nodes are connected together to form a circle, they represent an ${\rm SU}(2)$ vector multiplet along with a chiral multiplet with R-charge $2/3$ in the adjoint of that ${\rm SU}(2$). A class-$\mathcal{S}$ theory of genus~$g$ arises when $2g-2$ of these blocks are glued back-to-back (and forth-to-forth) along a straight line, with the leftmost and the rightmost blocks having two of their half-circular nodes glued together; the gauge group~$G$ is then ${\rm SU}(2)^{3(g-1)}$. Fig.~\ref{fig:g3S} shows the genus three example, which has six nodes and thus $G={\rm SU}(2)^{6}$.

The Romelsberger index of the genus $g$ theory takes the form:
\begin{gather*}
\mathcal{I}_{\mathcal{S}_g}(b,\beta)=\frac{(p;p)^{3(g-1)}(q;q)^{3(g-1)}}{2^{3(g-1)}}\int
\mathrm{d}^{3(g-1)}x\, \Delta_{\mathrm{block}}(x_1,x_1,x_2)\Delta_{\mathrm{block}}(x_2,x_3,x_4)\\
\hphantom{\mathcal{I}_{\mathcal{S}_g}(b,\beta)=}{}\times \Delta_{\mathrm{block}}(x_3,x_4,x_5)\cdots
\Delta_{\mathrm{block}}(x_{3g-4},x_{3(g-1)},x_{3(g-1)}),\label{eq:classSindex}
\end{gather*}
where
\begin{gather*}
\Delta_{\mathrm{block}}(x_1,x_2,x_3):=\Gamma\big((pq)^{1/3}z_1^{\pm1} z_2^{\pm1} z_3^{\pm1}\big)\frac{\Gamma\big((pq)^{1/3}\big)^3\prod\limits_i
\Gamma\big((pq)^{1/3}z_i^{\pm2}\big)}{\prod\limits_i \Gamma(z_i^{\pm2})},
\end{gather*}
with $z_i=e^{2\pi i x_i}$.

An ${\rm SU}(2)$ vector multiplet along with its accompanying adjoint chiral multiplet contribute to the Rains function of the theory as
\begin{gather*}
L_h^{\mathrm{node}}(x)=-\frac{2}{3}\vartheta(2x).\label{eq:SSNRains}
\end{gather*}
A semi-circular node contributes half as much, and thus the three semi-circular nodes in Fig.~\ref{fig:classSb} contribute together as
\begin{gather}
L_h^{\text{three semi-nodes}}(x,y,z)=-\frac{1}{3}\left(\vartheta(2x)+\vartheta(2y)+\vartheta(2z)\right).\label{eq:3NRains}
\end{gather}

The eight chiral multiplets represented by the triangle in Fig.~\ref{fig:classSb} contribute to the Rains function of the theory as
\begin{gather}
L_h^{\mathrm{triangle}}(x,y,z)=\frac{1}{3}\left(\vartheta(x+y+z)+\vartheta(x+y-z)+\vartheta(x-y+z)+\vartheta(-x+y+z)\right).\!\!\!\label{eq:T2Rains}
\end{gather}

Adding up (\ref{eq:3NRains}) and (\ref{eq:T2Rains}) we obtain the contribution of a single block to the Rains function:
\begin{gather}
L_h^{\mathrm{block}}(x,y,z)=\frac{1}{3}[\vartheta(x+y+z)+\vartheta(x+y-z)+\vartheta(x-y+z)+\vartheta(-x+y+z)\nonumber\\
\hphantom{L_h^{\mathrm{block}}(x,y,z)=}{} -\vartheta(2x)-\vartheta(2y)-\vartheta(2z)].\label{eq:blockRains}
\end{gather}

With the Rains function of the block (\ref{eq:blockRains}) at hand, we can now write down the Rains function of genus $g$ class-$\mathcal{S}$ theories. For example, the Rains function of the $g=2$ theory is given by
\begin{gather*}
L_h^{\mathcal{S}_{g=2}}(x_1,x_2,x_3)=L_h^{\mathrm{block}}(x_1,x_1,x_2)+L_h^{\mathrm{block}}(x_2,x_3,x_3),\label{eq:Sg=2Rains}
\end{gather*}
and the Rains function of the $g=3$ theory (illustrated in Fig.~\ref{fig:g3S}) is obtained as
\begin{gather*}
L_h^{\mathcal{S}_{g=3}}(x_1,x_2,x_3,x_4,x_5,x_6)=L_h^{\mathrm{block}}(x_1,x_1,x_2)+L_h^{\mathrm{block}}(x_2,x_3,x_4)\nonumber\\
\hphantom{L_h^{\mathcal{S}_{g=3}}(x_1,x_2,x_3,x_4,x_5,x_6)=}{} +L_h^{\mathrm{block}}(x_3,x_4,x_5)+L_h^{\mathrm{block}}(x_5,x_6,x_6).
\end{gather*}

Importantly, Rains's GTI (\ref{eq:RainsGTI}), with $c_1=x+y$, $c_2=x-y$, $d_1=z$, $d_2=-z$, implies that
\begin{gather*}
L_h^{\mathrm{block}}(x,y,z)\ge0.
\end{gather*}
It is not difficult to show that the equality holds in a~finite-volume subspace of the $x$, $y$, $z$ space; take for instance $x,y,z\approx .1$ within~$.01$ of each other, and use the fact that for small argument~$L_h$ reduces to~$\tilde{L}_{S^3}$ to show that $L_h$ vanishes in the domain just described.

Since the Rains function of a $g\ge2$ class-$\mathcal{S}$ theory is the sum of several block Rains functions, the positive semi-definiteness of $L_h^{\mathrm{block}}$ guarantees the positive semi-definiteness of $L_h^{\mathcal{S}_{g\ge2}}(x_i)$; hence
\begin{gather*}
L^{\mathcal{S}_{g\ge2}}_{h,\min}=0.
\end{gather*}
Moreover, taking all $x_i$ to be around $0.1$, and within $0.01$ of each other, we can easily conclude (as in the previous paragraph) that for the genus $g$ theory
\begin{gather*}
\dim \mathfrak{h}^{\mathcal{S}_{g\ge2}}_{qu}=3(g-1).
\end{gather*}
These results first appeared in \cite{Ardehali:thesis}.

\subsection[The ${\rm SU}(2)$ ISS model]{The $\boldsymbol{{\rm SU}(2)}$ ISS model}\label{subsec:ff}

The Intriligator--Seiberg--Shenker (ISS) model is an ${\rm SU}(2)$ gauge theory with a single chiral multiplet of R-charge $3/5$ in the four-dimensional (or spin-$3/2$) representation of the gauge group.

The Romelsberger index of this theory is (cf.~\cite{Vartanov:2010})
\begin{gather}
\mathcal{I}_{\rm ISS}(b,\beta)=\frac{(p;p)(q;q)}{2}\int\mathrm{d}x \, \frac{\Gamma\big((pq)^{3/10}z^{\pm 1}\big)\Gamma\big((pq)^{3/10}z^{\pm 3}\big)}{\Gamma\big(z^{\pm 2}\big)}.\label{eq:ISSindex}
\end{gather}

\begin{figure}[t]\centering
 \includegraphics[scale=1.1]{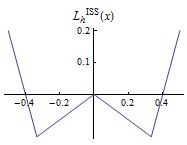}
\caption{The Rains function of the ${\rm SU}(2)$ ISS theory.}\label{fig:ISS}
\end{figure}

The Rains function of the theory is
\begin{gather*}
L^{\rm ISS}_h(x)=\frac{2}{5}\vartheta(x) +\frac{2}{5}\vartheta(3x)-\vartheta(2x).
\end{gather*}
This function is plotted in Fig.~\ref{fig:ISS}.

A direct examination reveals that $L^{\rm ISS}_h(x)$ is minimized at $x=\pm 1/3$, and $L^{\rm ISS}_h(\pm 1/3)=-2/15$. Therefore
\begin{gather*}
L^{\rm ISS}_{h,\min}=-2/15,\qquad \dim \mathfrak{h}^{N_c,N_f}_{qu}=0.
\end{gather*}

\subsubsection*{All-orders asymptotics}

A careful study shows \cite{Ardehali:2015c}
\begin{gather*}
\log \mathcal{I}_{\rm ISS}(b,\beta)\sim -\frac{\pi^2}{3\beta}\left(\frac{b+b^{-1}}{2}\right)\big(\operatorname{Tr}R+12L^{\rm ISS}_{h,\min}\big)
+ \log Y^{\rm ISS}_{S^3}(b)+\beta E_{\mathrm{susy}}(b)\quad \text{as} \ \beta\to 0\label{eq:ISSindexAsy3}
\end{gather*}
with
\begin{gather}
Y^{\rm ISS}_{S^3}(b)=\int_{-\infty}^{\infty}\mathrm{d}x'\,
e^{-\frac{4\pi}{5}(b+b^{-1})x'} \Gamma_h(3x'+(3/5)\omega)\Gamma_h(-3x'+(3/5)\omega),\label{eq:Yiss}
\end{gather}
and $\operatorname{Tr}R=7/5$. A numerical evaluation using
\begin{gather*}
\log \Gamma_h(ix;i,i)=(x-1)\log \big(1-e^{-2\pi i x}\big)-\frac{1}{2\pi i}\Li_2\big(e^{-2\pi i x}\big)+\frac{i\pi}{2}(x-1)^2-\frac{i\pi}{12},
\end{gather*}
yields $Y^{\rm ISS}_{S^3}(b=1)\approx 0.423$.

The EHI in (\ref{eq:ISSindex}) gives the simplest example known to the author where $L_h$ is minimized away from the origin. The hyperbolic reduction (\ref{eq:Yiss}) is unusual, compared with the cases studied by Rains \cite{Rains:2009} for which the hyperbolic reduction has essentially the same integrand as the elliptic integral but with hyperbolic gamma functions replacing elliptic gamma functions.

The content of this subsection first appeared in~\cite{Ardehali:2015c}.

\section{Open problems of physical interest}\label{sec:open}

We take the relation (\ref{eq:LagEquivIndex}) as our definition of an EHI. Let us first restrict attention to non-chiral EHIs: those in which the non-zero weights $\rho^j$ come in pairs with opposite signs. (All the EHIs studied by Rains in \cite{Rains:2009} are non-chiral.)

As mentioned below (\ref{eq:LagIndexSimp1}), the range of integration in an EHI can be interpreted in the gauge theory picture as the classical moduli space $\mathfrak{h}_{\rm cl}$ of holonomies around the circle $S^1_\beta$ of the background spacetime. The function $V^{\mathrm{eff}}$, and hence also the Rains function~$L_h$ (defined in~(\ref{eq:LhDef}), and related to $V^{\mathrm{eff}}$ as in (\ref{eq:VeffEquiv1})), can then be interpreted as a quantum effective potential for the holonomies. The locus of minima of $L_h$ gives the ``quantum moduli space'' $\mathfrak{h}_{qu}$ of the holonomies.

According to our main result (\ref{eq:LagIndexSimp6noTheta}), when the quantum moduli space $\mathfrak{h}_{qu}$ consists only of the origin $\boldsymbol{x}=0$ (where $L_h$ always vanishes) the hyperbolic asymptotics of the EHI is given by the Cardy-like formula of~\cite{DiPietro:2014}:
\begin{gather*}
\log\mathcal I(b,\beta)= -\frac{\pi^2}{3\beta}\left(\frac{b+b^{-1}}{2}\right)\operatorname{Tr}R+O\big(\beta^0\big),
\end{gather*}
with $\operatorname{Tr}R$ defined in~(\ref{eq:TrR}). The hyperbolic reduction of such EHIs works as in the cases studied by Rains~\cite{Rains:2009}.\footnote{Since~$Q_h$ is stationary at $\boldsymbol{x}=0$, it seems like when $\mathfrak{h}_{qu}=\{\boldsymbol{x}=0\}$ the analysis of the hyperbolic asymptotics should proceed similarly for chiral theories with non-zero~$Q_h$ as well.} An example of this relatively simple scenario is provided by the EHI $I^{(m)}_{A_n}$ of the ${\rm SU}(N_c)$ SQCD theory, analyzed in Section~\ref{subsec:sqcd}.

For EHIs whose quantum moduli space $\mathfrak{h}_{qu}$ is extended, but still contains the origin $\boldsymbol{x}=0$, relation~(\ref{eq:LagIndexSimp6noTheta}) gives the hyperbolic asymptotics as
\begin{gather*}
\log\mathcal I(b,\beta)= -\frac{\pi^2}{3\beta}\left(\frac{b+b^{-1}}{2}\right)\operatorname{Tr}R+\dim \mathfrak{h}_{qu}\log\left(\frac{2\pi}{\beta}\right)+O\big(\beta^0\big).
\end{gather*}
For such EHIs the hyperbolic reduction is more subtle than in the cases studied by Rains~\cite{Rains:2009}. The EHI~(\ref{eq:SONindex}) of the ${\rm SO}(2N+1)$ SQCD theory analyzed in Section~\ref{subsec:soN} is an example realizing this scenario. More examples of this kind can be found in \cite[Section~3.2]{Ardehali:2015c}. The EHIs of the class-$\mathcal{S}$ theories, discussed in Section~\ref{subsec:classS} above, are also examples of this kind.

Now, the most physically interesting scenario is when $L_h$ is minimized away from the origin, in which case the hyperbolic asymptotics looks like
\begin{gather*}
\log \mathcal{I}(b,\beta)= -\frac{\pi^2}{3\beta}\left(\frac{b+b^{-1}}{2}\right)(\operatorname{Tr}R+12L_{h,\min})+\dim \mathfrak{h}_{qu}\log\left(\frac{2\pi}{\beta}\right)+O\big(\beta^0\big),
\end{gather*}
and the hyperbolic reduction of the EHI is again more subtle than in the cases studied in~\cite{Rains:2009}.

In this last scenario, we can make an analogy with the Higgs mechanism in the standard model of particle physics. The ISS model of Section~\ref{subsec:ff} gives a clear example. Its Rains function resembles a \emph{Mexican-hat potential} familiar from the Higgs mechanism. We might roughly say that in this example an ``infinite-temperature Higgs mechanism'' moves the quantum moduli space $\mathfrak{h}_{qu}$ away from the origin (``infinite-temperature'' because the hyperbolic limit is roughly like the high-temperature limit in the gauge theory picture). In particle physics, Higgs mechanism describes the spontaneous breaking of gauge groups. Analogously we see an ${\rm SU}(2)\to{\rm U}(1)$ breaking of the gauge group as we go from the ISS model's EHI (\ref{eq:ISSindex}) in the form of an ${\rm SU}(2)$ matrix-integral, to its hyperbolic reduction~(\ref{eq:Yiss}) which is roughly in the form of a ${\rm U}(1)$ matrix-integral.

Besides the ISS model, one other SUSY gauge theory with $L_h$ minimized away from the origin was studied in \cite{Ardehali:2015c} (more examples have since been studied in \cite{DiPietro:2017,Hwang:2018}); that is the BCI model (cf.~\cite{Vartanov:2010}), which has $G= {\rm SO}(n)$ with $n>1$, and a~single chiral multiplet of R-charge $4/(n+2)$ in the two-index symmetric traceless tensor representation of the gauge group. The result of~\cite{Ardehali:2015c} on the hyperbolic reduction for $n=5$ shows an ${\rm SO}(5)\to {\rm U}(1)\times {\rm SO}(3)$ breaking in that case.

For both the ISS and the BCI model, we have $\operatorname{Tr}R>0$: specifically $\operatorname{Tr}R_{\rm ISS}=7/5$ and $\operatorname{Tr}R_{BCI}=n-1$. Therefore the following problem arises~\cite{Ardehali:2015c,DiPietro:2017}.
\begin{Problem} Prove $($or disprove$)$ that, in all non-chiral EHIs, $L_{h,\min}<0$ only if $\operatorname{Tr}R>0$.
\end{Problem}
Note that in the statement of the above problem we are writing ``only if''; writing ``if and only if'' would be incorrect because the class-$\mathcal{S}$ theories of Section~\ref{subsec:classS} have $\operatorname{Tr}R=2(g-1)/3>0$ but $L_{h,\min}=0$.

Our case-by-case study also shows that in SUSY gauge theories whose $L_h$ is positive in a punctured neighborhood of $\boldsymbol{x}=0$, $L_h$ is positive semi-definite everywhere, implying (in combination with $L_h(\boldsymbol{x}=0)=0$) that $L_{h,\min}=0$. This brings up another problem.
\begin{Problem} Prove $($or disprove$)$ that, for non-chiral EHIs, if the function $L_{h}$ $($and thus $\tilde{L}_{S^3}$ as defined in~\eqref{eq:hypURains}$)$ is strictly positive in some punctured neighborhood of the origin, then $L_{h}$ is positive semi-definite.
\end{Problem}
The significance of the above problem arises from the fact that if $\tilde{L}_{S^3}$ is strictly positive in some punctured neighborhood of the origin, then the squashed three-sphere partition func\-tion~$Z_{S^3}(b)$ of the dimensionally reduced theory is finite~\cite{Ardehali:2015c}. Therefore in such cases no infinity obviously threatens the simplistic physical intuition (spelled out in the introduction) for the hyperbolic reduction of the Romelsberger index. On the other hand, in the ISS and BCI models, the function $Z_{S^3}(b)$ diverges, signalling the breakdown of the simplistic physical intuition and the need for an infinite-temperature Higgs mechanism to save the day. Whether an infinite-temperature Higgs mechanism can happen even when $Z_{S^3}(b)$ is finite, is the question formalized by the above problem.

Finally, the obvious problem of finding the hyperbolic asymptotics of \emph{chiral} EHIs remains open.
\begin{Problem} Find the hyperbolic asymptotics of the master EHI in \eqref{eq:LagEquivIndex}, without assuming it to be non-chiral. In particular, prove $($or disprove$)$ Conjecture~{\rm \ref{conjecture}}.
\end{Problem}

\appendix

\section[Continuous non-R symmetries: flavor fugacities and R-charge deformations]{Continuous non-R symmetries: flavor fugacities\\ and R-charge deformations}

In this appendix we explain how additional parameters, known as flavor fugacities, can be incorporated into the EHIs of SUSY gauge theories that have (compact, Lie) symmetries besides their U(1)$_R$. Moreover, we show that when these extra symmetries include U(1) factors, the R-charge assignment of the chiral multiplets can be continuously deformed. These are well-known matters in the community of physicists working on EHIs.

\subsection{U(1) flavor symmetries and R-charge deformation}

We say a SUSY gauge theory with U(1) R-symmetry (as in Definition~\ref{definition1}) has a U(1)$_a$ flavor symmetry, if to each of its chiral multiplets $\{\mathcal{R}_j,r_j\}$ we can assign a U(1)$_a$ \emph{flavor charge} $q_j\in\mathbb{R}$, such that the following anomaly cancellation condition holds:
\begin{gather*}
\sum_j q_j\sum_{\rho^j\in\Delta_{j}}\rho^j_l\rho^j_m=0\qquad \text{for all $l$, $m$}.
\end{gather*}
(A special case of such U(1) flavor symmetries is what in the physics literature is called a~``baryonic'' U(1) symmetry.)

Then the following two results are easy to establish.
\begin{itemize}\itemsep=0pt
\item A \emph{flavor fugacity} $u_a=e^{2\pi i\beta m_a}$ [wherein we take $m_a\in\mathbb{R}$] can be incorporated into the expression (\ref{eq:LagEquivIndex}), by simply modifying its numerator gamma functions to $\Gamma\big((pq)^{r_j/2} z^{\rho^j}u_a^{q_j}\big)$; the resulting function $\mathcal{I}(b,\beta,m_a)$ is continuous over $b,\beta,m_a\in{}]0,\infty[$, but not necessarily real for $m_a\neq0$. As an example, see the EHI in \cite[equation~(9.2)]{Spirido:2009}, in the special case where $s_i=(pq)^{\frac{1}{2}-\frac{N}{(K+1)N_f}}u_a$ and $t_i=(pq)^{-\frac{1}{2}+\frac{N}{(K+1)N_f}}u_a$.
\item The \emph{deformed R-charges} $r'_j=r_j+\lambda q_j$, for $\lambda(\in\mathbb{R})$ small enough such that $r'_j\in{}]0,2[$, can replace $r_j$ and lead to new SUSY gauge theories and new EHIs.
\end{itemize}

Generalization to several U(1) flavor symmetries is straightforward.

The ``non-uniqueness'' of R-charges in the presence of U(1) flavor symmetries raises the following question regarding Problem~1 in Section~\ref{sec:open}: should $\operatorname{Tr}R$ be strictly positive for the whole family labeled by $\lambda$(s)? Reference~\cite{DiPietro:2017} conjectures that $L_{h,\min}<0$ only if \smash{$\operatorname{Tr}R_\ast>0$}, with $\ast$ referring to the specific value of~$\lambda$(s) for which $a:=3\operatorname{Tr}R^3-\operatorname{Tr}R$ is maximized. (The R-symmetry chosen through this ``a-maximization'' procedure can potentially serve as the preferred U(1)$_R$ of the superconformal field theory that the SUSY gauge theory flows to in the infrared~\cite{Intriligator:2003jj}.)

\subsection{Semi-simple flavor symmetries}

Let $F$ be a semi-simple matrix Lie group of rank $r_F$. We say a~SUSY gauge theory with U(1) R-symmetry (as in Definition~\ref{definition1}) has flavor symmetry group $F$ if its chiral multiplets come in irreducible finite dimensional representations $\mathcal{R}^{F}_1,\mathcal{R}^{F}_2,\dots$ of $F$, such that the following anomaly cancellation condition holds:
\begin{gather*}
\sum_j \rho^{F,j}_l \sum_{\rho^j\in\Delta_{j}}\rho^j_m\rho^j_n=0\qquad \text{for all $l$, $m$, $n$}.
\end{gather*}
The weight $\rho^{F,j}$, assigned to the chiral multiplet $\{\mathcal{R}_j,r_j\}$, belongs to $\Delta^F$, the set of all the weights of the representation $\mathcal{R}^{F}_k$ in which the chiral multiplet sits.

Then \emph{flavor fugacities} $u_{F,1}=e^{2\pi i\beta m_{F,1}},\dots,u_{F,r_F}=e^{2\pi i\beta m_{F,r_F}}$ (wherein we take \smash{$m_{F,i}\in\mathbb{R}$}) can be incorporated into the expression~(\ref{eq:LagEquivIndex}), by simply modifying its numerator gamma functions to $\Gamma\big((pq)^{r_j/2} z^{\rho^j}u_F^{\rho^{F,j}}\big)$, where $u_F^{\rho^{F,j}}$ stands for $u_{F,1}^{\rho^{F,j}_1}\times\dots\times u_{F,r_F}^{\rho^{F,j}_{r_F}}$; the resulting function $\mathcal{I}(b,\beta,m_{F,1},\dots,m_{F,r_F})$ is continuous over $b,\beta,m_{F,1},\dots,m_{F,r_F}\in{}]0,\infty[$, but not necessarily real when some of $m_{F,i}$ are nonzero. As an example, see the EHI in equation~(4.3) of~\cite{Spirido:2009}, but use $u_F=(s_1,\dots,s_{N_f-1},t_1,\dots t_{N_f-1})$ instead of~$s_i$,~$t_i$.

Generalization to compact $F$, possibly containing several U(1) flavor symmetries besides a~semi-simple factor, is straightforward.

\subsection*{Acknowledgements}

I would like to thank P.~Miller and E.~Rains for discussions on the mathematical side, as well as G.~Festuccia, J.T.~Liu, and P.~Szepietowski for discussions on the physical side of the subject. I~am also grateful to the anonymous referees, whose informative comments and constructive feedbacks on a draft of this manuscript has contributed significantly to its subsequent improvement. This work was supported in part by the National Elites Foundation of Iran.

\pdfbookmark[1]{References}{ref}
\LastPageEnding

\end{document}